\documentclass[traditabstract]{aa}  

\usepackage{graphicx}
\usepackage{subfig}
\usepackage{longtable,lscape}
\usepackage{txfonts}
\usepackage{hyperref}
\hypersetup{colorlinks=true,citecolor=blue,linkcolor=blue,urlcolor=blue}
\begin{document}

\title{Evolution of the cluster optical galaxy luminosity function in
  the CFHTLS: breaking the degeneracy between mass and
  redshift. \thanks{Based on observations obtained with
    MegaPrime/MegaCam, a joint project of CFHT and CEA/IRFU, at the
    Canada-France-Hawaii Telescope (CFHT) which is operated by the
    National Research Council (NRC) of Canada, the Institut National
    des Sciences de l'Univers of the Centre National de la Recherche
    Scientifique (CNRS) of France, and the University of Hawaii. This
    work is based in part on data products produced at Terapix
    available at the Canadian Astronomy Data Centre as part of the
    Canada-France-Hawaii Telescope Legacy Survey, a collaborative
    project of NRC and CNRS. } }
  
\titlerunning{GLF evolution in the CFHTLS}
\authorrunning{Sarron et al.}

   \author{F. Sarron\inst{1},
           N. Martinet\inst{2},
           F. Durret\inst{1},
           C. Adami\inst{3}
 \offprints{Florian Sarron,  \email{sarron@iap.fr}}
}

   \institute{UPMC Universit\'e Paris 06, UMR 7095, Institut d'Astrophysique de Paris, 
                98bis Bd Arago, 75014 Paris, France\\
              \email{florian.sarron@iap.fr}
         \and
             Argelander-Institut f\"ur Astronomie, Universit\"at Bonn,
  Auf dem H\"ugel 71, D-53121 Bonn, Germany
         \and
          Aix Marseille Universit\'e, CNRS, LAM, Laboratoire d'Astrophysique de
  Marseille, Marseille, France
             }

   \date{Received 25 September 2017 / Accepted 15 January 2018}

 
  \abstract
{Obtaining large samples of galaxy clusters is important for
  cosmology:  cluster counts as a function of redshift and mass
  can constrain the parameters of our Universe. They are also useful
  in order to understand the formation and evolution of clusters. We develop an
  improved version of the Adami \& MAzure Cluster FInder (AMACFI), now  the  Adami, MAzure \& Sarron Cluster FInder (AMASCFI), and
  apply it to the $154~{\rm deg}^2$ of the Canada-France-Hawaii
  Telescope Legacy Survey (CFHTLS) to obtain a large catalogue of 1371
  cluster candidates with mass $M_{200}>10^{14}\ {\rm M_\odot}$ and
  redshift $z\leq0.7$. We derive the selection function of the
  algorithm from the Millennium simulation, and cluster masses from a
  richness--mass scaling relation built from matching our candidates
  with X-ray detections.  We study the evolution of these clusters
  with mass and redshift by computing the $i'$-band galaxy luminosity
  functions (GLFs) for the early-type (ETGs) and late-type galaxies
  (LTGs). This sample is 90\% pure and 70\% complete, and therefore our
  results are representative of a large fraction of the cluster
  population in these redshift and mass ranges. We find an increase in
  both the ETG and LTG faint populations with decreasing redshift
  (with Schechter slopes $\alpha_{\rm ETG}=-0.65\pm0.03$ and
  $\alpha_{\rm LTG}=-0.95\pm0.04$ at $z=0.6$, and $\alpha_{\rm
    ETG}=-0.79\pm0.02$ and $\alpha_{\rm LTG}=-1.26\pm0.03$ at $z=0.2$)
  and also a decrease in the LTG (but not  the ETG) bright end. Our
  large sample allows us to break the degeneracy between mass and
  redshift, finding that the redshift evolution is more pronounced in
  high-mass clusters, but that there is no significant dependence of
  the faint end on mass for a given redshift. These results show that
  the cluster red sequence is mainly formed at redshift $z>0.7$, and
  that faint ETGs continue to enrich the red sequence through
  quenching of brighter LTGs at $z\leq0.7$. The efficiency of this
  quenching is higher in large-mass clusters, while the accretion rate of faint LTGs is lower as the more massive clusters have already emptied most of their environment at higher redshifts.}


   \maketitle
%

\defcitealias{M07}{M07}
\defcitealias{A10}{A10}
\defcitealias{D11}{D11}
\defcitealias{M15}{M15}
\defcitealias{Leauthaud10}{L10}
\defcitealias{Kettula15}{K15}

\section{Introduction}

As the most massive gravitationally bound structures in the universe, clusters of galaxies have been observed in great detail for decades. In addition to  being interesting astrophysical objects, they are also a powerful probe of cosmology since galaxy cluster counts as a function of mass and redshift depend on the cosmological parameters of our Universe  \citep[see e.g.][for a review]{Allen+11}. 

In this context it is important to obtain extensive samples of clusters covering wide redshift and mass ranges. It is also necessary to know the selection function of the sample with great precision, as the cosmological constraints are obtained by comparing the observed cluster counts to the predicted ones, either from the analytical halo mass function or from N-body simulations. The development of extended imaging surveys, such as the Canada-France-Hawaii Telescope Legacy Survey (CFHTLS), provided the community with large sets of galaxy clusters observed homogeneously. It has therefore become possible to detect thousands of galaxy clusters up to redshifts $z\sim 1$ \citep[e.g.][]{A10, D11}. The large sky coverage of these surveys demands an automated detection, and cluster detection algorithms are a hot topic in the literature, both for the present and next generation surveys, such as that foreseen with \textit{Euclid}\footnote{\url{http://www.euclid-ec.org}}, which will uncover hundreds of thousands of clusters. \\
\indent Many detection algorithms exist with differences in the selection function focusing either on purity or completeness. These algorithms can be separated into parametric and non-parametric. In the first category are matched filtering algorithms \citep[e.g.][]{Bellagamba+17}, which apply a filter (e.g. Gaussian smoothing, Schechter function) to the galaxy field to highlight clusters, and red-sequence algorithms \citep{RedGold}, which use colour cuts to detect the linear relation between colour and magnitude of early-type cluster galaxies. Among the non-parametric are friend-of-friend algorithms \citep{Farrens+11}, which match close galaxies  with a characteristic scale, or wavelet algorithms \citep{Eisenhardt+08}, which use the information from different scales through wavelet transformations. In the present paper we developed an improved version of the Adami \& MAzure Cluster FInder \citep[AMACFI,][]{M07} algorithm which applies a smoothing to the galaxy density field in photo-$z$ slices.

Galaxy clusters can also be used to study galaxy evolution through the distribution of galaxy magnitudes (i.e. the galaxy luminosity function, hereafter GLF) of different galaxy types. In particular the evolution of the faint end of the GLF with redshift and mass gives insights into the effect of environment on the quenching of star formation. For nearby clusters, the faint end of red passive galaxies appears to be flat \citep[e.g.][]{Gaidos97,Paolillo+01}, while it experiences a mild decrease with redshift \citep[e.g. ][]{Smail+98,DeLucia04,Tanaka05,DeLucia07,Stott+07,Gilbank+08,Rudnick+09,Vulcani+11,M15,Zenteno+16,Martinet+17}. We note that some authors find no evolution with redshift \citep[e.g. ][]{Andreon08,DePropris+07,DePropris+13}. Recently \citet{Martinet+17} ruled out the possibility that these differences could arise from surface brightness selection biases between ground- and space-based observations. Therefore, differences in the GLF faint end cannot be due to observing conditions, but are more likely to reflect variations from one cluster to another in samples of typically a few tens of clusters. The dependence of the GLF on mass is less studied, mainly because of the degeneracy between mass and redshift. Indeed, at higher redshift we only detect the most massive clusters, while at low redshift we have a complete sample in mass. Attempts to break this degeneracy have failed so far because of the high number of clusters  required \citep{M15,Martinet+17}.

In the present study, we have improved the AMACFI cluster detection algorithm, and renamed this new version AMASCFI (Adami, MAzure \& Sarron Cluster FInder). We apply it to the 154 square degrees covered by the four Wide fields of the CFHTLS survey. In particular, cluster positions and redshifts are now more accurate than the previous version used by \citet{D11}. We detect a total of  3743 cluster candidates up to redshift $z\sim 1.1$ and estimate the selection function of this sample by applying the same algorithm to numerical simulations based on lightcones from the Millennium simulation by \citet{Springel05} and modified by \citet{Henriques12}. We derive a mass for each cluster candidate using a richness--mass relation calibrated on the X-ray clusters of \citet{Gozaliasl} and \citet{Mirkazemi15} that are also detected by our algorithm. We compute cluster GLFs for all galaxies, and also for early and late types separately. We study the evolution of the faint end with redshift and mass independently. And finally, we make use of the high number of clusters to break the degeneracy between mass and redshift for the first time. 

The paper is structured as follows. In Sect.~\ref{sec:detection} we
describe the AMASCFI code and its recent improvements.  In
Sect.~\ref{selection_function} we study the properties of AMASCFI using simulations. In Sect.~\ref{sec:AMASCFI_T0007} we compare our catalogue  of cluster candidates to the literature using optically and X-ray detected clusters.  In
Sect.~\ref{GLFsection} we explain how we derived the GLFs, and the
corresponding results are given in Sect.~\ref{GLF_results}.  The
results are discussed in Sect.~\ref{sec:discussion} and the conclusions
are drawn in Sect.~\ref{sec:conclusion}.
We use AB magnitudes throughout the paper, and assume a flat
$\Lambda$CDM cosmology with $\Omega_{\rm M}=0.3$ and $h=0.7$.

\section{Cluster detection} \label{sec:detection}
We have updated the Adami \& MAzure Cluster FInder \citep[AMACFI,][]{M07} and applied it to the CFHTLS final data release T0007 photometric redshift (hereafter photo-$z$, symbol $z_{\rm phot}$) catalogues. The original AMACFI algorithm was already applied to the CFHTLS in previous studies: \citet{M07} for the Deep1 field, \citet{A10} for the T0004 data realease, and \citet{D11} for the Wide fields of the T0006 data release. We briefly present the main features of the method, focusing on the improvements, their motivations, and their implications.

\subsection{Photometric redshift catalogue}
The photo-$z$ catalogue is obtained from the CFHTLS data release T0007\footnote{available at \url{http://terapix.iap.fr/article.php?id_article=841}}. 

\indent CFHTLS T0007 photo-$z$s were computed in the 154 deg$^{2}$ sky coverage of CFHTLS using multicolour images in the $u^{*}g'r'i'z'$ filters of MegaCam at CFHT. We note that the $i'$ filter had to be changed during the course of the survey. The photo-$z$s were obtained using the LePhare software \citep[]{LePhare99, Ilbert06}. \\
\indent Details about the method are given in \citet{Coupon09}. Briefly, the photo-$z$s were computed using 62 templates obtained after having optimized four templates from \citet{CWW} and two starburst templates from \citet{KinneySB}, and linearly interpolated between them to better sample the colour-redshift space using the VVDS spectroscopic sample \citep[e.g.][]{VVDS05}. A particularly crucial step of the process is the calibration of the zero-points using spectroscopic samples which help in removing biases. The resulting statistical errors on photo-$z$s depend on the redshifts and magnitudes of the galaxies.\\
\indent Following the photo-$z$ catalogue based on the CFHTLS T0007 data release, we define the dispersion as 
\begin{equation}
\sigma _{\Delta z_{\rm phot} / (1 + z_{\rm s})} = 1.48 \times \mathrm{median} \, \left ( \frac{\left | \Delta z  \right |}{(1+z_{\rm s})} \right), 
\end{equation}
\noindent which is the NMAD estimator defined in \citet{Ilbert06}, with $\Delta z_{\rm phot} = z_{\rm phot} - z_{\rm s}$, where $z_{\rm s}$ is the spectroscopic redshift. The outlier rate or catastrophic failure rate $\eta$ is set as the proportion of objects with $\left | \Delta z  \right | \geq 0.15 \times (1+z_{\rm s})$. \\
\indent We make use of the value reported in the release document to choose our cuts in redshift and magnitude. For cluster detection, we select galaxies in the redshift range $0.1 < z_{\rm phot} < 1.2$ and with magnitudes $i' < 22.5$, thus keeping the dispersion below $0.05 \times (1+z)$ and the outlier rate below $10 \%$ in all four Wide fields.

\indent In our analysis, we only consider galaxies that are outside the masks from TERAPIX. These masks are located around bright stars or artefacts, and mark regions of lower photometric quality. Thus photo-$z$s in these regions would be of poorer quality than those outside the masked regions. These masks are dealt with in the same way for cluster detection and GLF computation, i.e. by discarding objects inside the masked regions. 
We note that this approach is different from the one \citet{Moutard16} applied to the CFHTLenS, but we prefer to use the prescription from the TERAPIX team as we are using their photo-$z$ catalogue. 

\subsection{AMASCFI: description of the algorithm} \label{sec:AMACFI}

Our detection algorithm is based on the method described in \cite{M07}. The galaxy catalogue is cut into slices of redshift, partially overlapping so as not to miss structures. \cite{M07} and subsequent studies using AMACFI chose a constant slice width of 0.1 in redshift space, each slice overlapping the adjacent ones by 0.05. In contrast, we adopt here a variable width chosen as
\begin{eqnarray}
\Delta z_{\rm slice} &=& 0.05  \times (1+z_{\rm slice})\,, \\ 
z_{\rm slice} \,(n+1) &=& z_{\rm slice} \,(n) + 0.05 \,,
\end{eqnarray}

\noindent This enables us to better account for the noise due to photo-$z$ statistical errors, and therefore to sample a galaxy population representative of the true underlying population in each redshift slice, especially at high redshift. For example, the slice width at $z=1.1$ (maximum redshift considered in our analysis) is now taken to be $\Delta z_{\rm slice} = 0.2$ rather than 0.1. The galaxies were selected in the redshift range $0.1 < z < 1.2$, thus the first slice is  centred at $z=0.16$ and the last at $z=1.10$,  respectively the minimum and maximum redshifts of our detected cluster candidates. Even though the photo-$z$ dispersion is slightly different from field to field, and depends on the redshift and magnitudes of the sources, we decided to be conservative and consider it as $\sigma _{\Delta z_{\rm phot} / (1 + z_{\rm s})} = 0.05$ in all four fields, so we can treat them as homogeneously as possible. 

Two-dimensional density maps are then computed in each slice using an adaptive kernel density estimator (adaptive-KDE). So far, AMACFI chose the initial size of the kernel automatically according to the \citet{silverman1986density} prescriptions. We decided in this work to fix the initial kernel size to 1.5 Mpc.
   
\indent In this way, the kernel size (diameter) of the adaptive-KDE in the densest region (corresponding to galaxy clusters) is $\sim 1$ Mpc, which is the right smoothing scale for detecting clusters as it is the typical size of cluster cores. In fact, our previous way of choosing the smoothing scale in AMACFI was oversmoothing the underlying distribution and was thus more suited for supercluster than cluster detection. 

The SExtractor software is then applied to the density maps to detect structures. The major modification in this step of the process is the way the detection threshold is set. This threshold is now set to a number of galaxies per Mpc$^{2}$. 

For this, we iteratively compute the background level (field) in galaxies per Mpc$^{2}$ in each slice using our density map, and set the detection threshold to the 95\% upper confidence limit on this number density following \citet{Gehrels86}.

On the first iteration, the field is simply set to the mean galaxy number density in the entire density map. We compute the detection threshold and use SExtractor to detect overdensities in the map. The background level is updated as the mean galaxy number density in the map after having removed pixels of the density map in a disk of diameter 1 Mpc around the peak of each detected structure. This gives us a new detection threshold and the process is repeated until convergence. At the end of the procedure, we obtain the final field level in the slice $\langle n_{\rm field} \rangle$ , and the detection threshold thus quantifies the probability for an overdensity to be a random fluctuation of the background (due to chance alignment in the photo-$z$ space).

We compute the mean number density of galaxies in a disk of $1$ Mpc diameter centred at the peak of each overdensity as detected by SExtractor from our density maps and obtain a signal-to-noise ratio ($S/N$) of detection for each overdensity. The $S/N$ of detection is defined as
\begin{equation}
S/N = \frac{\langle n_{\rm clus}\rangle - \langle n_{\rm field} \rangle}{\sqrt{\langle n_{\rm field} \rangle}},
\end{equation}

\noindent where $\langle n_{\rm clus} \rangle$ and $\langle n_{\rm field} \rangle$ correspond to the average number density of galaxies per Mpc$^{-2}$ in a slice of width $\Delta z_{\rm slice}$ for cluster and field area, respectively.

The overdensities thus detected in each slice are then assembled in larger structures (called \textit{cluster candidates} in the following) using a friends-of-friends algorithm, the \textit{Minimal Spanning Tree}, with a characteristic distance of 1 Mpc, as in \cite{M07} \citep[see][for the original description of the algorithm]{MST99}. This allows us to merge multiple detections of the same structure appearing in several adjacent redshift slices, or of large clusters presenting many substructures. The position of each \textit{candidate cluster} in the (RA, Dec, $z_{\rm phot}$) space is taken to be the mean of each of its individual merged detections weighted by its excess galaxy number density ($\langle n_{\rm clus} \rangle - \langle n_{\rm field} \rangle$).

\section{AMASCFI selection function}\label{selection_function}
   \begin{figure*}
\centering
     \includegraphics[width=0.8 \textwidth]{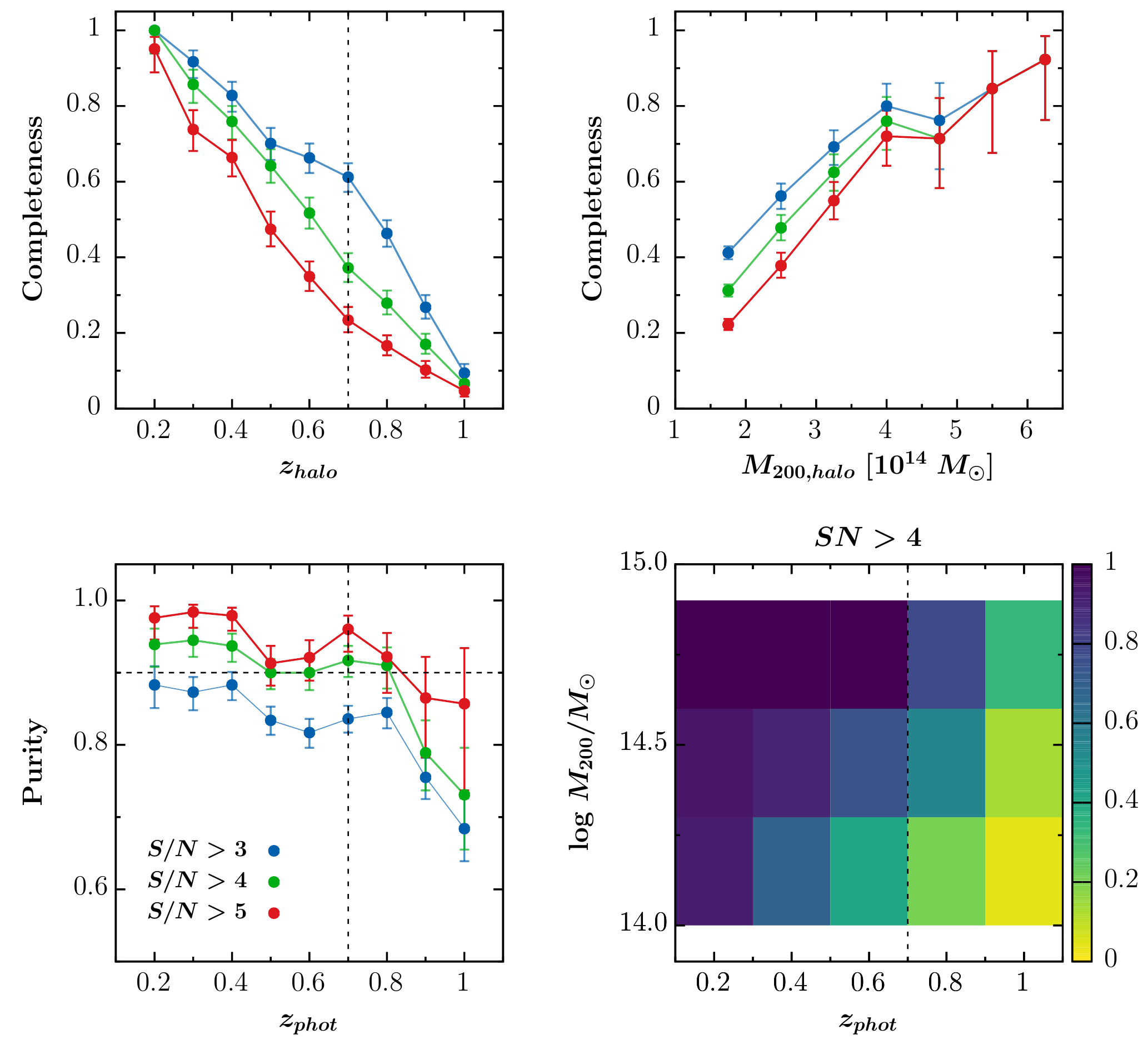}
     \caption{Selection function of AMASCFI using the Millennium modified lightcones of \citet{Henriques12}.  \textit{Top left}: Completeness as a function of redshift for haloes of $M_{200} > 10^{14} \ {\rm M_\odot}$. \textit{Top right}: Completeness as a function of halo mass. \textit{Bottom left}: Purity as computed for haloes of mass $M_{200} > 10^{13} \ {\rm M_\odot}$. In these three panels, blue, green, and red points respectively show the results for $S/N > 3$, $S/N > 4$, and $S/N > 5$.  \textit{Bottom right}: Two-dimensional histogram of the completeness in the (redshift, mass) parameter space for cluster candidates with $S/N > 4$. The vertical dotted line at $z=0.7$ shows the cut applied to compute GLFs (see Sect.~\ref{sec:GLFs_cuts}) and the horizontal line the 90\% purity limit.}
     \label{fig:detrate}
   \end{figure*}

In order to calibrate our method, i.e. to assess the reliability of our detections, we apply our algorithm to a set of 24 lightcones computed by \citet{Henriques12} from the Millennium Simulation \citep{Springel05} and built using the \citet{Guo11} semi-analytical model. Since then, a new set of lightcones has been built by \citet{Henriques15} using the Planck cosmology rather than the original Millennium Simulation cosmology (WMAP1). Even so,  when comparing galaxy number counts with CFHTLS data, the \citet{Henriques12} lightcones were in better agreement than the more recent ones. 
The total area of the 24 independent beams is $\sim50 \ {\rm deg}^{2}$, and thus contains $\sim1000$ haloes in the redshift range $0.1< z < 1.2$ with mass $M_{200} > 10^{14} \ {\rm M_\odot}$, where $M_{200}$ is the mass contained in a radius $r_{200}$,  inside which the density is 200 times the critical density of the universe. Such a cosmological volume enables us to properly assess the selection function of AMASCFI.
In the following, we present the modifications we applied to the lightcone to make it a fair representation of our data, and we compute the selection function of AMASCFI.

\subsection{Mock catalogue modification}

We converted the SDSS magnitudes of the simulated mocks to CFHTLS Megacam $i'$-band magnitudes following the relation from the Megacam pages\footnote{\url{http://www.cadc-ccda.hia-iha.nrc-cnrc.gc.ca/en/megapipe/docs/filtold.html}}:
\begin{equation}
i_{\mathrm{Megacam}} = i_{\mathrm{SDSS}} - 0.085 \times (r_{\mathrm{SDSS}}- i_{\mathrm{SDSS}}).
\end{equation}

\noindent To make the mock galaxy catalogues from \citet{Henriques12} comparable to our data, we need to add realistic noise to the redshift of each galaxy in the mock. Since the error on the photo-$z$s depends  on redshift and on magnitude, we compute the mean $1 \sigma$ uncertainty on individual photo-$z$s (as given by the LePhare software) in bins of 0.1 in redshift and 0.25 in magnitude for the W1 field. On average, this $1 \sigma$ uncertainty closely follows  the statistical error computed using spectroscopic redshifts, justifying the use of this quantity for our purpose. We decided not to add noise on the lightcone magnitudes themselves as the errors for the CFHTLS T0007 data release are below 0.01 for 95\% of the sample at $i' < 23$, and thus negligible compared to photo-$z$ errors. 

We then apply a Gaussian error with a zero mean and a standard deviation corresponding to the mean LePhare $1 \sigma$ uncertainty in the corresponding bin of the lightcones. As in \citet{A10}, we did not account for catastrophic errors on the photo-$z$s in this simplified model. The effect on cluster detection is expected to be small since the outlier rate stays well below 10\% for the chosen magnitudes and redshift cuts.

The final step consists in applying a masking procedure representative of the one used by TERAPIX on the CFHTLS T0007. To this end, we modify the sky coordinates of the galaxies in the lightcones to match a subarea of the CFHTLS representative of the observed masks and apply the \textit{VENICE} program\footnote{\url{http://jeancoupon.com/venice/}} to remove galaxies from the masked regions. This step is important because masking could have an impact on the detection level of a cluster whose centre falls near a mask boundary. 

We thus obtain lightcones resembling CFHTLS T0007 data in terms of masking and photometric redshift distribution, on which we can accurately compute our selection function.

\subsection{Completeness and purity of the cluster catalogue obtained with AMASCFI} \label{sec:sel_fun_Mil}

In  order to assess the quality of our cluster detection we need a way to quantify how well we detect actual overdensities of galaxies and how polluted by false detections our catalogue is. Following the literature, we compute the \textit{completeness (C)} and the \textit{purity (P)} of our catalogue of cluster candidates to study the performances of AMASCFI. These two quantities are widely used to infer the quality of cluster catalogues. They are defined as 
\begin{eqnarray}
C = N_{\rm match}/N_{\rm det},\\
P = N_{\rm match}/N_{\rm true},
\end{eqnarray}

\noindent where $N_{\rm det}$ is the number of cluster candidates detected in the simulation, $N_{\rm true}$ is the total number of haloes in the simulation, and $N_{\rm match}$ the number of detected clusters matched to a halo from the simulation.

Ideally,  an algorithm should have high completeness (all clusters are detected) and high purity (all detections are actual clusters), but  there is a tradeoff between the two quantities and both cannot be maximized  simultaneously. The quantities are both functions of the S/N cuts, which can vary depending on application.
In Sect. \ref{GLFsection}, for our study of GLFs, we wanted to choose a threshold that guarantees high purity so there is no contamination by false detections whose GLFs would resemble those of field galaxies.

We ran our detection algorithm on the simulation exactly in the same way as on the CFHTLS data, and thus obtain a catalogue of cluster candidates with a sky position, redshift, and detection significance. To compute the completeness and purity of the catalogue as a function of significance, redshift, and mass, we match our candidate cluster catalogue with the halo catalogue from the simulation. The centre of each halo is taken to be its central galaxy. To match a candidate cluster with a halo, we ranked our cluster candidates by significance and the halo catalogue by halo mass and consider that they are matched when  

$\bullet$ the sky projected distance of the centres is less than the radius $r_{200}$ of the halo at the halo redshift;

$\bullet$ $\Delta z = \left | \ z_{\rm halo} - z_{\rm AMASCFI} \ \right | \le 0.1 \times (1+z).$

Since we are interested in detecting clusters of galaxies, we compute our completeness for haloes of mass $M_{200} > 10^{14} \ {\rm M_\odot}$. However,  computing purity with the same mass threshold would be too drastic. Indeed, some of our detections may correspond to haloes of smaller masses (galaxy groups) and thus not be false detections {per se}. This arises from the intrinsic scatter ($\sigma_{\rm int}$) that exists between halo mass and cluster richness. This scatter is found to be of the order $0.3$ dex in the local Universe \citep{Andreon12}. This means a halo with mass $\sim 3 \sigma_{\rm int}$ below the  $10^{14} \ {\rm M_\odot}$ threshold could have the same richness as a halo of $10^{14} \ {\rm M_\odot}$. Following these considerations, we compute the purity of our candidate cluster catalogue for haloes of mass $M_{200} > 10^{13} \ {\rm M_\odot}$. 

\begin{figure}[!]
\centering
   \includegraphics[width=0.45\textwidth]{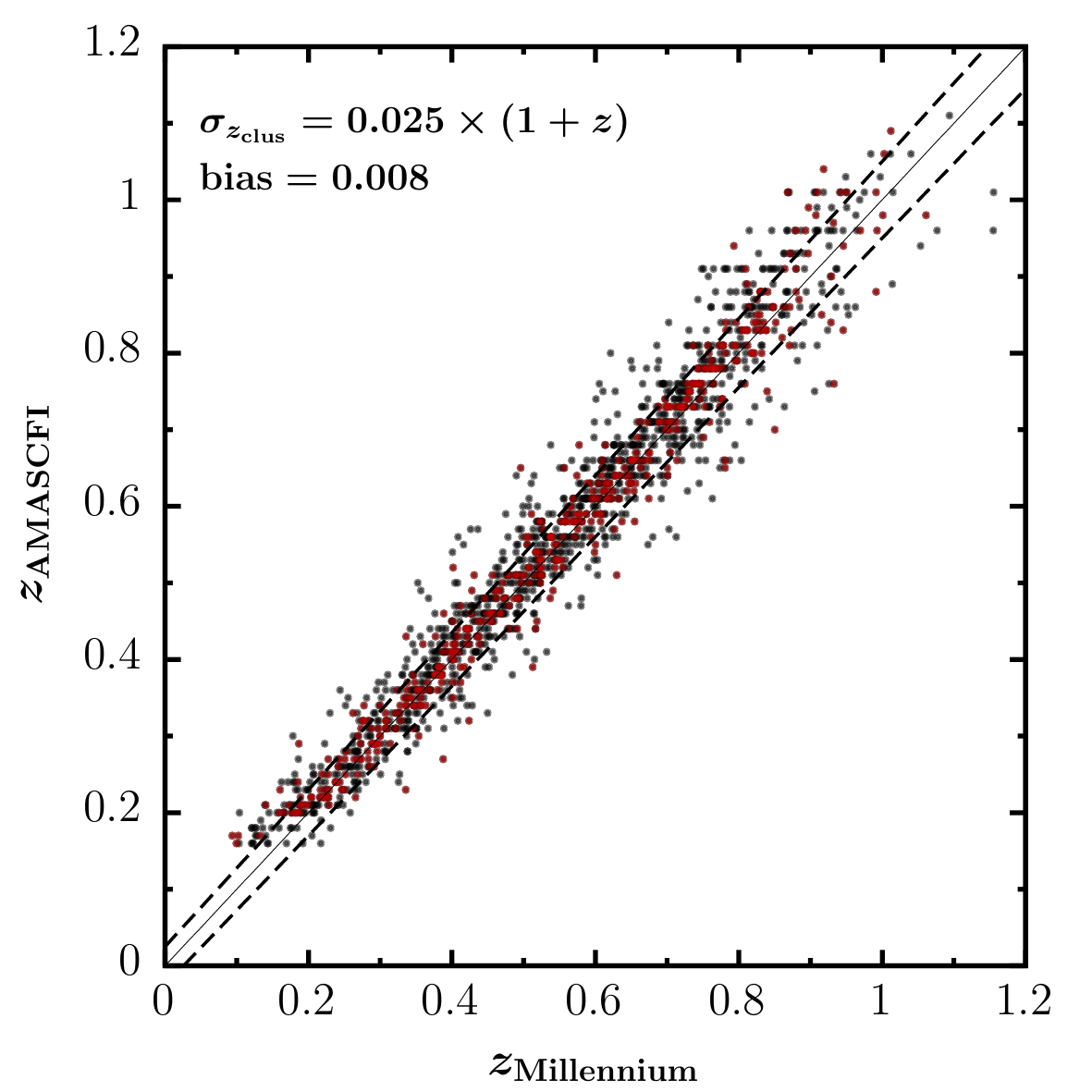}
     \caption{$z_{\rm AMACSFI}$ vs. $z_{\rm Millennium}$ for all cluster candidates with $S/N > 3$ matched in the \citet{Henriques12} lightcones. Black points are for all clusters and groups with mass $M_{200} > 10^{13} \ {\rm M_\odot}$, while red points are only for clusters with mass $M_{200} > 10^{14} \ {\rm M_\odot}$. The black line is the identity line. The dashed lines are $\pm \sigma_{z_{\rm clus}} = 0.025 \times (1+z)$.}
          \label{fig:zp_vs_zsim}
   \end{figure}

In Fig.~\ref{fig:detrate}, we show the cluster completeness as a function of redshift for haloes of mass $M_{200} > 10^{14} \ {\rm M_\odot}$ and as a function of halo mass in the entire redshift range $0.15 < z \le 1.1$, for different $S/N$ cuts. Error bars represent the $1 \sigma$ confidence limit of a binomial distribution following \citet{Gehrels86}. The bottom left panel of Fig.~\ref{fig:detrate} shows the purity of our catalogue, while the bottom right panel breaks the degeneracy between mass and redshift, showing the completeness as a function of both parameters for $S/N > 4$. The error on the redshift assigned to a candidate cluster by AMASCFI ($z_{AMASCFI}$) is estimated using the NMAD estimator. We find $\sigma_{z_{\rm clus}} = 0.025 \times (1+z)$ for haloes of mass $M_{200} > 10^{13} \ {\rm M_\odot}$. When considering only matched haloes of mass $M_{200} > 10^{14} \ {\rm M_\odot}$, this reduces to $\sigma_{z_{\rm clus}} = 0.025 \times (1+z)$. This is illustrated in Fig.~\ref{fig:zp_vs_zsim} where we plot $z_{AMASCFI}$ versus the true redshift of the matched halo in the simulation $z_{Millennium}$.

When computing completeness and purity as a function of redshift only, bins include clusters with a redshift $\pm 2 \times \sigma_{\rm zclus}$ from the central redshift of the bin, where $\sigma_{\rm zclus}=0.025 \times (1+z)$ is the statistical error on cluster redshift. The first bin is centred at $z=0.2$ and the last at $z=1.0$, with an offset of 0.1 between bins. When computing completeness as a function of mass, bins include clusters with a mass $\pm 7.5 \times 10^{13} \ {\rm M_\odot}$ from the central mass of the bin. The first bin is centred on $1.75 \times 10^{14} \ {\rm M_\odot}$ and consecutive bins are offset by $\pm 7.5 \times 10^{13} \ {\rm M_\odot}$, so that bins partially overlap and smooth the selection function. The last bin in mass only includes a lower limit and so includes all clusters more massive than $5 \times 10^{14} \ {\rm M_\odot}$.

When breaking the degeneracy between redshift and mass, we consider three bins in each parameter. This is done so that each bin is sufficiently populated to have reliable statistics. We choose to use the three redshift bins [0.1, 0.3[, [0.3, 0.5[, and [0.5, 0.7[ and mass bins ]$10^{14} \ {\rm M_\odot}$, $10^{14.3} \ {\rm M_\odot}$], ]$10^{14.3} \ {\rm M_\odot}$, $10^{14.6} \ {\rm M_\odot}$], and ]$10^{14.6} \ {\rm M_\odot}$, $\infty$ [. The mass bins were built to have the same size in logarithmic space and so that the highest mass--highest redshift bin contains a sufficiently large number of clusters.

For the three $S/N$ cuts considered, the purity is > 70\% in the entire redshift range, and >80\% for $z_{clus} \le 0.8$. The completeness is > 70\% for the most massive clusters ($\log{M_{200}/{\rm M_\odot}} > 14.6$) in the entire redshift range, while being always $\sim$70\% for all clusters ($\log{M_{200}/{\rm M_\odot}} > 14$) up to $z_{clus} =0.4$. As the redshift increases, large differences can be seen in the completeness depending on the $S/N$ cut considered. The completeness becomes low (< 30\%) for all $S/N$ cuts for $z_{clus} > 0.8$. This is due to the increasing errors on the photo-$z$s with redshift. 

Since we want to use our cluster catalogue to compute cluster GLFs, the primary criterion is a high purity of the catalogue. We thus choose a cut at $S/N > 4$ that guarantees a purity > 90\% up to $z_{clus} = 0.8$. For this very pure cut, we compute the completeness as a function of both redshift and halo mass (bottom right panel of Fig.~\ref{fig:detrate}). For the most massive bin ($\log{M_{200}/{\rm M_\odot}} > 14.6$), the completeness is > 70\% up to $z_{bin} = 0.8$ and then drops. A 70\% completeness threshold is reached up to $z_{bin} = 0.6$ for the intermediate mass bin ($14.3 < \log{M_{200}/{\rm M_\odot}} \le 14.6$) and $z_{bin} = 0.4$ for the lowest mass bin ($14 < \log{M_{200}/{\rm M_\odot}} \le 14.3$). In addition to its high purity, the cluster candidate catalogue used for the GLF computation is therefore also representative of clusters with mass $M_{200} > 10^{14} \ {\rm M_\sun}$.

\begin{figure*}[h]
\centering
     \includegraphics[width=1.\textwidth]{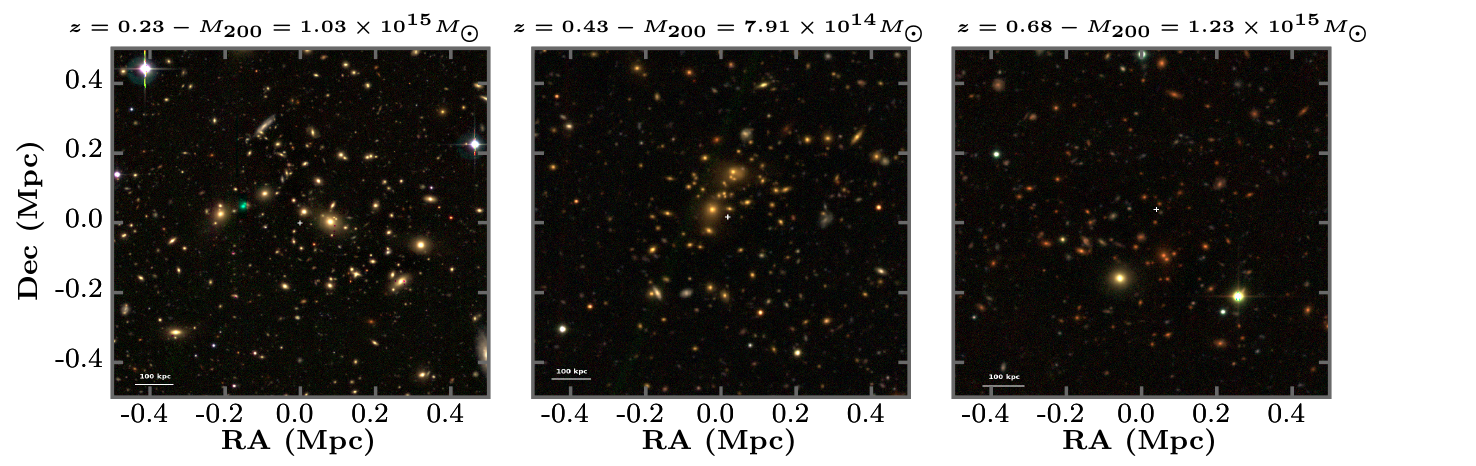}
     \caption{{\it gri} images of three rich cluster candidates in the CFTHLS W1 field, centred on the AMASCFI cluster centres.}
     \label{fig:RS}
   \end{figure*}
\subsection{Mass, redshift, and $S/N$ cuts for GLFs} \label{sec:GLFs_cuts}
        
In the following sections of the paper, we study the properties of cluster GLFs, using the cluster candidates detected with AMASCFI. A study of this kind is usually done with a catalogue of confirmed clusters because the results are relatively sensitive to contamination by false detections. Thus, to study the cluster candidate GLFs, we need to choose cuts in S/N, redshift, and richness (or equivalently mass), which ensures a high purity. At the same time, we would like to keep the completeness to high enough levels so that the population of cluster candidates under consideration can be considered as fairly representative of the true cluster population.

Using the selection function we computed on simulations, we chose to cut our \textit{final catalogue} for GLF computation at $S/N>4$, $z<0.7$ and $M_{200} > 10^{14} \ {\rm M_\odot}$. This ensures a purity higher than 90\% in the full redshift range (see bottom left panel of Fig.~\ref{fig:detrate}) and a completeness greater than 50\% in all the redshift and mass bins considered. The lower mass limit enables us to probe low-mass galaxy clusters, and the redshift range to study the cosmic evolution of the cluster properties.

\section{AMASCFI applied to the CFHTLS T0007 data release} \label{sec:AMASCFI_T0007}

When applying AMASCFI to the four Wide fields of the CFHTLS T0007 data release photo-$z$ catalogue, we detect 7100 cluster candidates at $S/N > 3$ in the redshift range $0.15 < z < 1.1$. The full catalogue will be made available at the CDS\footnote{\url{http://vizier.u-strasbg.fr/viz-bin/VizieR}}. Optical images of three rich clusters are shown in Fig.~\ref{fig:RS} for illustration. In the following sections, we will compare our candidate cluster catalogue to previously published cluster catalogues on the Wide fields of CFHTLS. We will first make a comparison with other optically detected cluster candidate catalogues from \citet{RedGold} and \citet{Ford+15}, and with X-ray detected cluster catalogues from \citet{Gozaliasl} and \citet{Mirkazemi15}.

\subsection{Matching AMASCFI cluster candidates with other optically selected cluster candidates} \label{sec:Ford_Licitra}

There are two public catalogues of optically selected cluster candidates, obtained from the CFHTLS observations,  although both use the CFHTLenS photometric catalogue extracted from the data rather than the TERAPIX catalogue.
The first is the \citet{Ford+15} cluster catalogue. It covers the four Wide fields of the CFHTLS and was obtained using the 3D-MF algorithm developed by \citet{Milkeraitis+10}. The 3D-MF algorithm is a matched filter algorithm that assumes a cluster radial profile and luminosity function and detects clusters in overlapping slices of redshift. The full published catalogue contains 22694 cluster candidates with significance $\sigma_{\rm Ford} > 3.5$ in the redshift range $0.2 < z < 1.0$. \\
\indent To match the two catalogues we use the same matching procedure as in Sect.~\ref{sec:sel_fun_Mil}, except for the maximum radial projection between the centres of two matched clusters, which is fixed to 2 Mpc to account for the large errors on the sky position of cluster candidates and because we have no estimate of $r_{200}$ in the observations. When matching our catalogue with the full 3D-MF catalogue, we find 5285 cluster candidates in common, meaning that 75\% of AMASCFI clusters have a counterpart in the 3D-MF catalogue. This agrees well with the purity of AMASCFI at $S/N > 3$, that we computed using the Millennium simulation (see Sect.~\ref{sec:sel_fun_Mil}), considering the 3D-MF cluster catalogue has high completeness \citep{Milkeraitis+10}. If we trim the 3D-MF catalogue to $\sigma_{\rm Ford} > 5$ (which corresponds to $\sim 1.5 \times 10^{13} \ {\rm M_\odot}$)  and $\sigma_{Ford} > 10$ (which corresponds to $\sim 10^{14} \ {\rm M_\odot}$; \citealt{Ford+15}),  there are respectively 6544 and 282 clusters left in the redshift range $0.2 < z < 1.0$. AMASCFI respectively detects 3521 ($\sim54\%$) and 260 ($\sim92\%$) of these. \\
\indent \citet{RedGold} developed the RedGOLD cluster detection algorithm, based on the search for red-sequence galaxy overdensities. The search is done in slices of redshift, where the red-sequence colour is predicted using stellar population models. To select their cluster candidates, they impose a Navarro--Frenk--White (NFW) profile, and compute a richness estimator $\lambda_{\rm RedGOLD}$ for each candidate. RedGOLD was applied to the CFHTLS W1 field. Their published catalogue includes 652 cluster candidates in the redshift range $0.14 \le z < 1.2$. Out of the 7100 cluster candidates detected by AMASCFI, 2951 lie in the CFHTLS W1 field.\\
\indent We use the same matching procedure as for the \citet{Ford+15} catalogue. Out of the 652 RedGOLD cluster candidates, 510 are also found by AMASCFI ($78\%$), a result in good agreement with the $\sim 80 \%$ purity of both catalogues at the significance cuts used. If we only consider AMASCFI cluster candidates with $S/N > 5.5$, to have a comparable number of candidates in both catalogues (663 for AMASCFI and 652 for RedGold), $45\%$ of cluster candidates are matched. Since the RedGOLD catalogue has an announced completeness of $\sim70\%$ and AMASCFI has less than $50\%$ (at this $S/N$ cut) in the redshift range considered, this is expected.\\

\subsection{Matching AMASCFI cluster candidates with X-ray detected clusters} \label{sec:Xray}
        
We  also compare our candidate cluster catalogue with two X-ray detected cluster catalogues provided by \citet{Gozaliasl} and \citet{Mirkazemi15}, both obtained from \textit{XXM-Newton} observations. The \citet{Gozaliasl} catalogue covers an area of 3 deg$^{2}$ inside the CFHTLS W1 field and includes 135 X-ray detected groups and clusters up to $z =1.1$, while the \citet{Mirkazemi15} catalogue was built by pointing at given optically selected cluster candidates;  it includes 196 X-ray detected groups and clusters up to $z =1.1$ in the CFHTLS W1, W2, and W4.

\indent Both catalogues provide $M_{200}$ for each cluster, obtained by applying the scaling relation between weak lensing mass and X-ray luminosity  ($M_{200,WL}-L_{X}$) obtained by \citet{Leauthaud10}. We recalibrate these masses according to the \citet{Kettula15} scaling relation as it presents the advantage of measuring the weak lensing mass for individual clusters, while \citet{Leauthaud10} stacked the lowest mass clusters in quite poorly populated bins.

\indent The intrinsic scatter in the relation and the errors on the fit parameters are propagated to obtain the errors on the mass estimates. Finally, we translated cluster masses into our own cosmology (only $H_{0} = 70 \ \mathrm{km~s^{-1}~Mpc^{-1}}$ differs).

\indent When comparing the mass distributions obtained from the $M_{200,WL}-L_{X}$ relations to that obtained by applying AMASCFI to the Millennium simulation, we find that the \citet{Leauthaud10} $M_{200,WL}-L_{X}$ relation underpredicts the number of clusters with $M_{200} > 10^{14} \ {\rm M_\odot}$, while the number of clusters predicted by the \citet{Kettula15} relation is in good agreement. \citet{Parroni17} recently found similar results, with the \citet{Leauthaud10} normalization being too low, when compared to their CFHTLenS data, especially in the range $M_{200} > 10^{14} \ {\rm M_\odot}$,  which is of main interest in our study. These arguments lead us to believe that our choice to use \citet{Kettula15} $M_{200,WL}-L_{X}$ scaling relation in our analysis was the correct one.

\indent Having obtained these mass estimates, we can match our cluster candidates with the X-ray detected clusters and look how well AMACSFI redetects them as a function of mass and redshift. However, neither \citet{Gozaliasl} nor \citet{Mirkazemi15} provide the completeness of their catalogue for given mass and redshift. Thus they cannot be used to properly compute a selection function of our algorithm, but only give us insights into how AMASCFI performed on the CFHTLS T0007 data. We use the same matching procedure as in Sect.~\ref{sec:sel_fun_Mil}, using for $r_{200}$ the value given in the original catalogue. 

\begin{figure}[h]
\centering
     \includegraphics[width=0.4\textwidth]{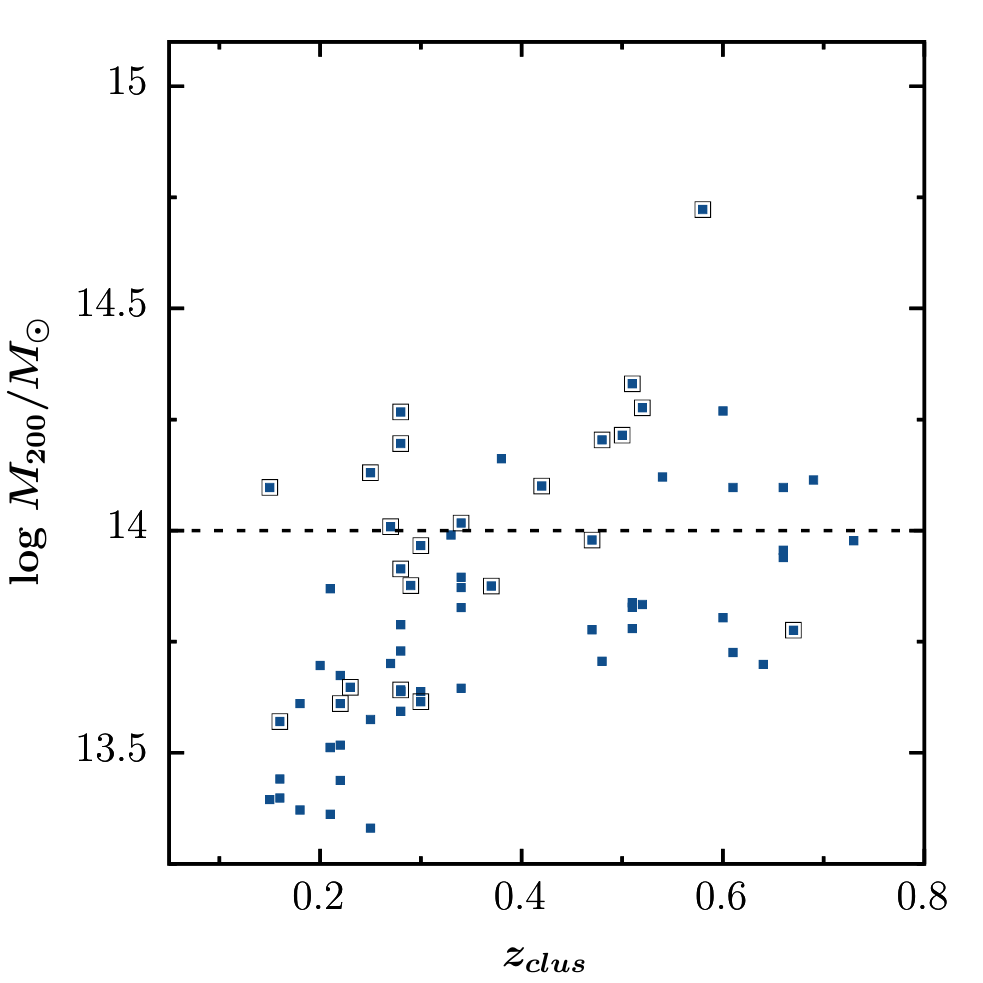}
     \caption{Matching of our cluster candidates obtained with AMASCI and X-ray groups detected in the same area of the W1 field by \citet{Gozaliasl}. Blue filled squares are clusters and groups from the \citet{Gozaliasl} catalogue. Black empty squares are clusters also detected by AMASCFI. See text for details.}
     \label{fig:detrate_XGoz}
   \end{figure}

\begin{figure}[h]
\centering
     \includegraphics[width=0.4\textwidth]{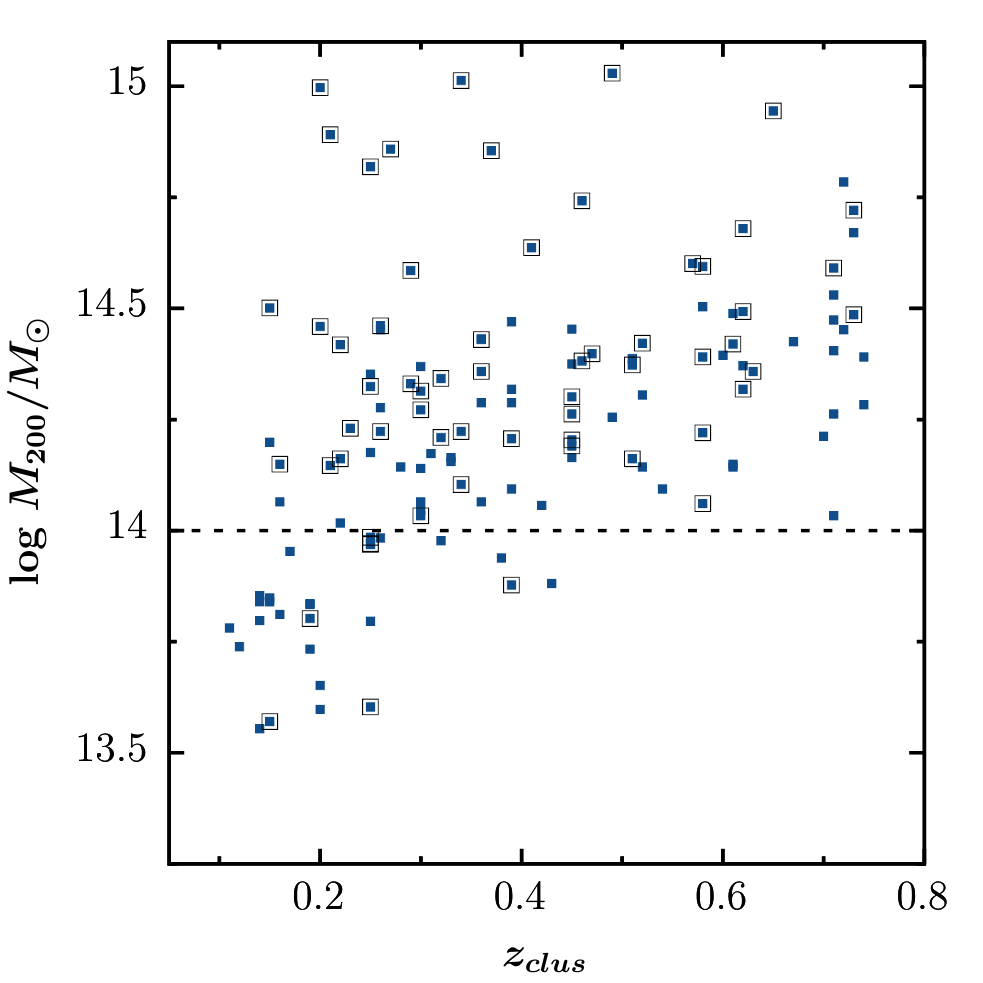}
     \caption{Matching of our cluster candidates obtained with AMASCFI and X-ray groups/clusters detected in the W1, W2, and W4 fields by \citet{Mirkazemi15}. The symbols are the same as in Fig.~\ref{fig:detrate_XGoz}.}
     \label{fig:detrate_XMir}
   \end{figure}

There are 68 groups in the \citet{Gozaliasl} catalogue with $z < 0.75$. In the same area AMASCFI detects 51 clusters at $S/N > 4$ and $z < 0.7$, 23 of them also being in the \citet{Gozaliasl} catalogue (see Fig.~\ref{fig:detrate_XGoz}). In addition, AMASCFI detects all but two X-ray clusters with $M_{200} > 10^{14} \ {\rm M_\odot}$ up to $z = 0.6$. The \citet{Mirkazemi15} catalogue contains 130 clusters with $z < 0.75$, while AMASCFI detects 1872 clusters at $S/N > 4$ and $z < 0.7$ in the W1, W2, and W4 fields. There are 66 clusters in common between the two catalogues (see Fig.~\ref{fig:detrate_XMir}). When using the detections in common with the X-ray catalogues, the statistical error on the redshifts of the AMASCFI clusters is the same as that computed on the Millennium simulation ($\sigma_{\rm clus} = 0.025$).

We did not match our cluster candidates with X-ray detected clusters from XXL because the published catalogue from \citet{Pacaud+16} only includes massive clusters and thus covers a mass range less interesting than the \citet{Gozaliasl} and \citet{Mirkazemi15} catalogues combined.

\subsection{Mass-richness calibration} \label{sec:richness}

The  richness of a cluster is known to be a proxy for its mass, which is not measurable directly. Having an estimator with as small a scatter as possible in the mass-richness  relation is of great interest to study the dependence of cluster properties on mass. We can use our GLF computation method (see Sect.~\ref{GLFsection}) to derive such a richness estimator for our cluster candidates. \citet{Rykoff12lambda} showed that including blue cluster members in their richness estimate increased scatter in the $L_{X}$-richness relation from $\sigma_{\ln_{L_{X}} | \lambda} = 0.63$ to $\sigma_{ln_{L_{X}} | \lambda} = 0.72$ at the $2 \sigma$ level. With this result in mind, we decided to build our richness estimator based on the cluster GLF of early-type galaxies (ETGs). Our goal was to count the number of red ETGs brighter than a given absolute magnitude, so we can directly use the counts in absolute magnitude that are also used to build our GLF. 
   \begin{figure}[h]
\centering
     \includegraphics[width=0.4\textwidth]{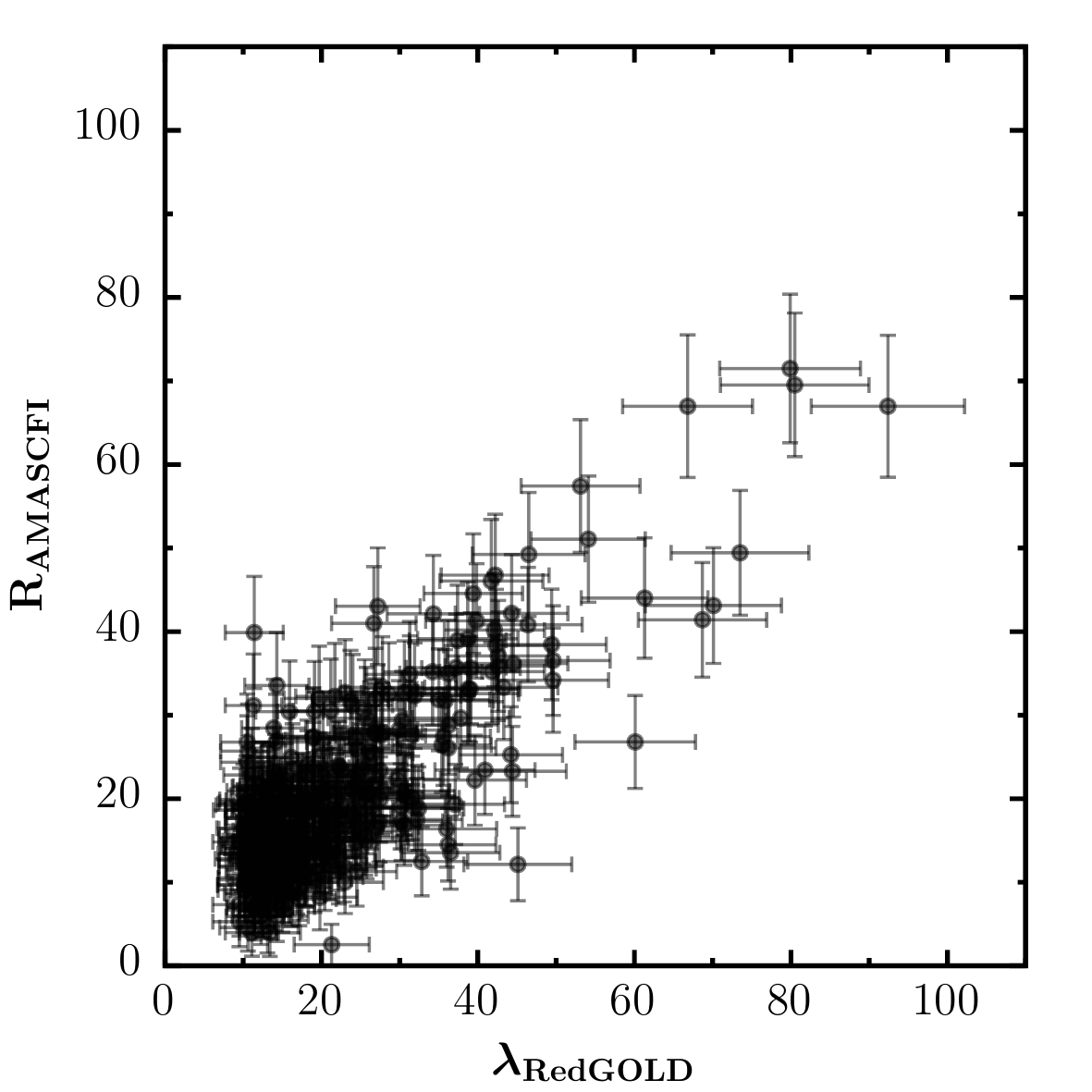}
     \caption{Comparison of the AMACFI (${\rm R_{AMASCFI}}$) and RedGOLD ($\lambda_{\rm RedGOLD}$) richness estimates.}
     \label{fig:richcomp}
   \end{figure}
   
We computed the ETG GLF in a 1 Mpc radius. We then summed the number counts in absolute magnitude bins, after removing field counts, for galaxies brighter than $0.2 \times L^{*}$, i.e. $M < M^{*} + 1.75$. Here $M^{*}$ (respectively $L^{*}$) is the characteristic absolute magnitude (luminosity) of a cluster, and is obtained from a Schechter fit to the GLFs stacked in redshift bins. We took a constant $M^*$ equal to the mean $M^*$ over our redshift range: $\langle M^* \rangle_{z} = -22.6$.

The 90\% completeness limit of our sample being $i'=23$, we were able to reach $M^*+1.75$ up to $z = 0.7$. Thus, our richness estimator was homogeneous up to $z=0.7$, and we did not compute the richness for clusters with higher redshifts. 

We calibrated our richness $R$ to X-ray derived cluster masses, using the catalogues from \citet{Gozaliasl} and \citet{Mirkazemi15} to infer the mass for all our cluster candidates up to $z=0.7$. We used the catalogue of matched clusters derived in Sect.~\ref{sec:Xray} and fitted the $M_{200}-R$ relation as
\begin{equation}
\mathrm{log_{10}} \ \frac{M_{200}}{{\rm M_\odot}} = \beta + \alpha \times \mathrm{log_{10}} \ \frac{R}{40},
\end{equation}

\noindent where the pivot value for richness is taken to be $40$  as in, e.g. \citet{RedGold,Rykoff12lambda}.

\begin{figure}[h]
\centering
     \includegraphics[width=0.45\textwidth]{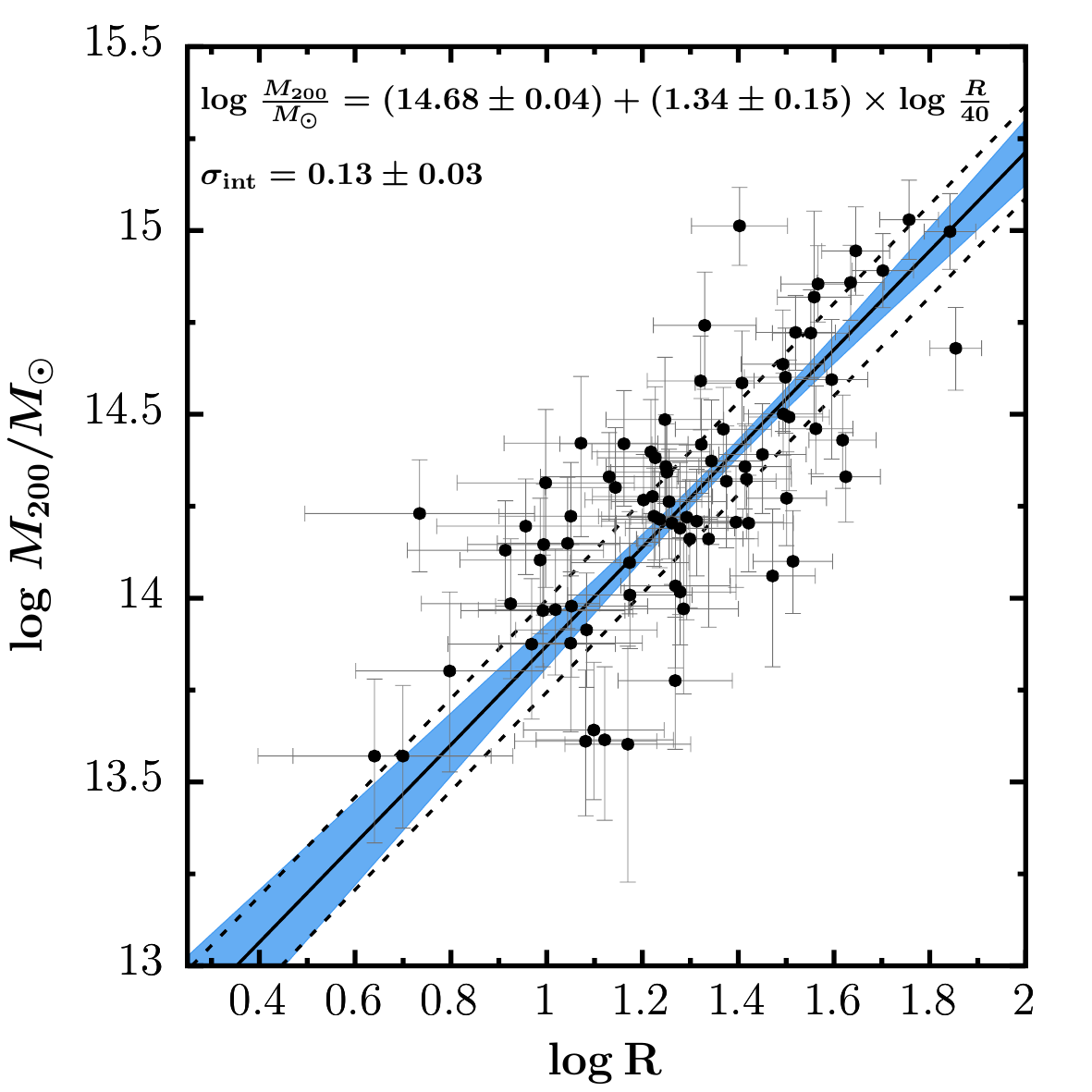}
     \caption{Mass-richness relation for clusters in common between AMASCFI (at $S/N > 4$ and $z < 0.7$) and either the \citet{Gozaliasl} or \citet{Mirkazemi15} catalogues. The mass $M_{200}$ of the X-ray detected clusters was obtained applying the \citet{Kettula15} $M_{200}-L_{X}$ scaling relation. The fitted relation is shown at the top of the figure. The solid black line is the median relation, while the blue zone shows the 68\% confidence on the fit parameters. The dashed lines show the intrinsic scatter in the relation.}
     \label{fig:mass-R}
   \end{figure}

Figure~\ref{fig:mass-R} shows the mass-richness relation for the 82 clusters detected by AMASCFI with an X-ray counterpart in either the \citet{Gozaliasl} or \citet{Mirkazemi15} catalogues up to $z_{AMASCFI}=0.7$. The fit was done using the John Meyers python implementation of the linmix\_err IDL routine \citep{Kelly07} with a default superposition of three Gaussians. This Bayesian approach considers that the errors  follow a Gaussian distribution, while the error on the richness is Poissonian by nature. However, we verified that every cluster has a high enough richness (R > 10) for the Poisson law to be closely approximated by a Gaussian. We note that four clusters only reach R > 5; we kept them in the analysis as they should not significantly affect the errors on the fit and improve the representation of low-mass clusters. We obtain $\beta = 14.68 \pm 0.04$, $\alpha = 1.34 \pm 0.15$, and the intrinsic dispersion $\sigma_{\rm int} = 0.13 \pm 0.03$. 

Comparing our scaling relation with the literature is not straightforward since different richness estimators will yield different scaling relations. However, our definition of richness is similar to that computed by the RedGOLD algorithm \citep{RedGold}, i.e. counting bright ETGs in a given radius. The main difference is that contrary to \citet{RedGold}, who scale the radius in which the richness is computed iteratively, we fixed this radius to 1 Mpc. We compare the two estimators in Fig.~\ref{fig:richcomp}, plotting the AMASCFI richness estimator ${\rm R_{AMASCFI}}$ versus the RedGOLD richness estimator $\lambda_{\rm RedGOLD}$ for the matched clusters. We find a good correlation between them, the AMASCFI richness being 16 \% lower in the mean than RedGOLD richness. 

\citet{Parroni17} used the RedGOLD catalogue to compute a mass-richness scaling relation from weak-lensing masses  and their results are similar to ours, but with lower $\alpha$ and $\beta$. This is expected because our richness estimates are lower than theirs. We also find a similar intrinsic scatter $\sigma_{\rm int}$, showing that both estimators are similarly good mass proxies, as previously argued by \citet{AndreonandHurn10}.

\indent Using the fit result, we can infer the posterior probability distribution for the mass of each candidate cluster for which we have computed a richness. As discussed previously (Sect.~\ref{sec:Xray}) this gives us a mass distribution for our cluster catalogue that is in good agreement with that derived from applying AMASCFI to simulations. 

\section{Cluster galaxy luminosity functions} \label{GLFsection}

We wanted to  study the evolution of the cluster luminosity functions (GLFs) with redshift and their dependence on cluster mass. Our method for computing GLFs is based on the method developed in \citet{M15} (hereafter \citetalias{M15}) and adapted to the specificity of CFHTLS T0007 data. Taking advantage of the size of our sample, we were able to break the degeneracy between mass and redshift. We computed GLFs of our cluster candidates in the $i'$ rest frame band using photo-$z$ information to estimate the cluster membership of galaxies. We used the $i'$ band because for photo-$z$ computation, the full T0007 catalogue was cut at $i'<24$, so it is complete in this band only. \citet{M15} showed that GLFs behave similarly in the V, R, and I band, so our conclusions are quite general.

\subsection{Completeness limit}

One key point when computing GLFs is to properly define the 90\% completeness limit of the sample. The CFHTLS data have the advantage of being homogeneous across the whole field and the 80\% completeness limits were computed by TERAPIX. However, these depth limits show substantial variations from \textit{tile} to \textit{tile}. Indeed, the deepest \textit{tile} has an  80\% completeness limit of $c_{i',80\%} = 24.07$, while the shallowest has $c_{i,80\%}' = 23.30$ (the mean being $\overline{c}_{i',80\%} = 23.72$). To obtain the 90\% completeness limit, we used  data provided in T0007 of the CFHTLS, where the completeness limit is assessed using simulations. 

We decided to be conservative and to take the same 90\% completeness limit in all the fields: $c_{i',90\%} = 23.0$.  This has two main advantages: the first is that we were able to study homogeneously  our entire sample, and the second is that photo-$z$s become noisier for magnitudes fainter than $i'=23$ ($\sigma_{\Delta z_{phot}} / (1 + z_s) \sim 0.07$ and $\eta > \sim 10-15\%$), so that considering fainter galaxies may have affected the quality of our analysis.

This apparent magnitude completeness limit is translated to an absolute magnitude completeness limit by adding the distance modulus and k-correction. We used the k-corrections computed by LePhare. The software computes the theoretical k-correction from the best fit template and best estimated redshift of the galaxy. Here we want our completeness value to be correct for all types of galaxies. We thus selected all the galaxies in $0.05 \times (1+z_{\rm clus})$ and computed for each template the mean k-correction. For our result to hold for both early- and late-type galaxies (ETGs and LTGs, see Sect.~\ref{sect:ETG_LTG}), we computed the mean k-correction over ETG templates and LTG templates and the final value was taken as the maximum of these two quantities. 

\subsection{Galaxy luminosity function computation} \label{GLF}

To compute our GLFs, we used the final catalogue containing all the relevant information for each galaxy: position in the sky, photo-$z$, and apparent and absolute magnitudes in $i'$ band. For each cluster, we selected galaxies in a cylinder of radius 1 Mpc and length $\pm \, 2 \times 0.05 \times (1+z_{\rm clus})$. Part of these galaxies are not cluster members but rather background or foreground galaxies, so we  needed to remove the field contribution because the photo-$z$ statistical error of $\sim 0.05 \times (1+z_{s})$ is larger than the typical size of a cluster $\sim 0.001 \times (1+z)$ \citep[see e.g.][]{Evrard08}. \\
We call the galaxies thus selected the \textit{candidate cluster galaxies} in the following. Once this selection was done, we fixed all the \textit{candidate cluster galaxies} to the cluster redshift and re-ran LePhare without fitting the photo-$z$s (parameter ZFIX). LePhare only fits the best template, which might be different when we force $z_{\rm gal} = z_{\rm clus}$ as discussed in \citetalias{M15}. This allows us to determine the template, k-correction, and hence the absolute magnitude of each \textit{candidate cluster galaxy} more accurately, as these are redshift dependent properties. 

We then subtracted the field contribution from our \textit{candidate cluster galaxies}. The field galaxies are defined for each redshift slice as galaxies more than 2 Mpc away from any detected cluster in the slice. We normalised the field to each cluster area before subtracting galaxy counts.

One point that needs to be dealt with carefully is  that field galaxies have k-corrections computed at their own redshift, while for the \textit{candidate cluster galaxies} the k-correction was computed at the cluster redshift. An error in the k-correction could move a galaxy from one absolute magnitude bin to another in our GLF, thus distorting the actual GLF. To avoid this effect, we removed the field contribution in bins of apparent magnitude. We counted \textit{candidate cluster galaxies} and \textit{field galaxies} in bins of 0.5 magnitude and applied a weight to all galaxies in the bin equal to the ratio of cluster to field galaxies in the bin. Once field counts were subtracted, we normalised our GLFs to 1 Mpc$^{2}$. This was done so that we can compare the GLFs at different redshifts.

\subsection{Stacking the GLFs} \label{GLF:stackmethod}

Because of low number counts, individual cluster GLFs are noisy. Thus, we cannot use them to infer the dependence of the faint-end slope of the GLF with mass and redshift. To increase our S/N, we stacked our GLFs in bins of redshift and mass. Stacking was done as in \citetalias{M15} using the standard Colless method \citep{Colless89}. The idea is to average cluster galaxy counts in each absolute magnitude bin, including all clusters that are 90\% complete in this bin. Clusters first have to be normalized to the same area and to a fixed richness. We normalised all clusters to 1 Mpc$^{2}$. For the richness we used our estimator described in Sect.~\ref{sec:richness}.

The main advantage of this method compared to a classical average of GLFs is that we are able to use as much information as possible. Indeed, with a classical method, the average would only be done in the absolute magnitude bins which are 90\% complete for all the clusters considered, thus limiting our capacity to probe the faint-end in a given redshift range. 

   \begin{figure*}[h]
\centering
   \includegraphics[width=1.\textwidth]{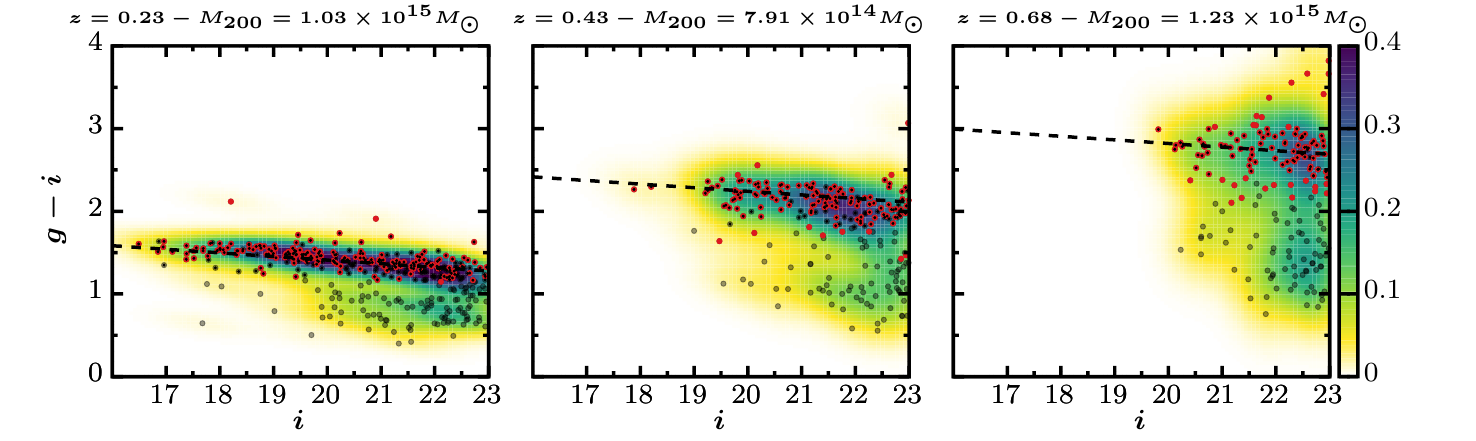}
     \caption{Colour-magnitude diagram ({\it g-i}) vs. {\it i} for three rich clusters in the CFHTLS W1 field. Empty black circles are candidate cluster galaxies. The underlying distribution is their normalized density. In red are galaxies selected as ETGs at the cluster redshift by \textit{LePhare}. Smaller black points are galaxies selected as RS members, i.e. lying $\pm 0.3$ from the best fit RS.}
     \label{fig:RS_vs_Lephare}
   \end{figure*}

Galaxy counts and their errors are summed following Eqs. \ref{eqn:Nstack} and \ref{eqn:sigstack}, respectively, where $N(j)$ and $\sigma (j)$ are the stacked galaxy counts and galaxy count errors in magnitude bins $j$, the index $i$ indicates single cluster values, $S_{i}$ is the area of cluster $i$, $N_{c}(j)$ is the number of clusters in the bin $j$, and $N_{0,i}$ and $\langle N_{0}(j) \rangle $ are the richness of cluster i and the mean richness of clusters in the $j$ bin: 

\begin{eqnarray}
N(j) = \frac{\langle N_{0}(j) \rangle }{N_{c}(j)} \sum_{i} \frac{N_{i}(j)}{S_{i}N_{0,i}}, \label{eqn:Nstack}\\
\sigma(j) = \frac{\langle N_{0}(j) \rangle} {N_{c}(j)} \sqrt{ \sum_{i} \left ( \frac{\sigma_{i}(j)}{S_{i}N_{0,i}} \right )^{2}} \label{eqn:sigstack}
\end{eqnarray}

\noindent To retain the Poissonian distribution of the counts, we weight the individual variances by the square of the cluster area, as for the galaxy counts, and not simply the area. We did not take into account the clustering error in our estimation of the individual variances, because it is negligible compared to the Poisson error. We fit the stacked $i'$ band GLFs with a Schechter function \citep{Schech76} 

 \begin{equation}
N(M) = 0.4 \, \mathrm{ln}(10) \, \phi^{*} \left [ 10^{0.4(M^{*}-M)} \right ]^{\alpha+1} \, e^{-10^{0.4(M^{*}-M)}},
\end{equation}

\noindent where $\phi^{*}$ is the characteristic number of galaxies per unit volume, $M^{*}$ the characteristic absolute magnitude, and $\alpha$ the faint-end slope of the GLF. The fit is done with a $\chi^{2}$ minimization. The error bars on the parameters correspond to the $1 \sigma$ confidence level and are computed from the covariance matrix, evaluated at the best parameter values. These single parameter error bars include the effects of correlations with other parameters. As in \citet{Martinet+17}, we convert the final $\chi^{2}$ value in a confidence probability $p$ assuming a $\chi^{2}$ distribution with three degrees of freedom ($\alpha$, M$^{*}$, $\phi^{*}$) : 
 \begin{equation}
p(\chi^{2},3) = \frac{2}{\sqrt{\pi}} \left [ \frac{\sqrt{\pi}}{2} \mathrm{erf} \left (\sqrt{\frac{\chi^{2}}{2}} \right ) - \mathrm{exp} \left ( - \frac{\chi^{2}}{2} \right ) \sqrt{\frac{\chi^{2}}{2}} \right ].
\label{eq:p}
\end{equation}

\begin{figure}[!]
\centering
   \includegraphics[width=0.4\textwidth]{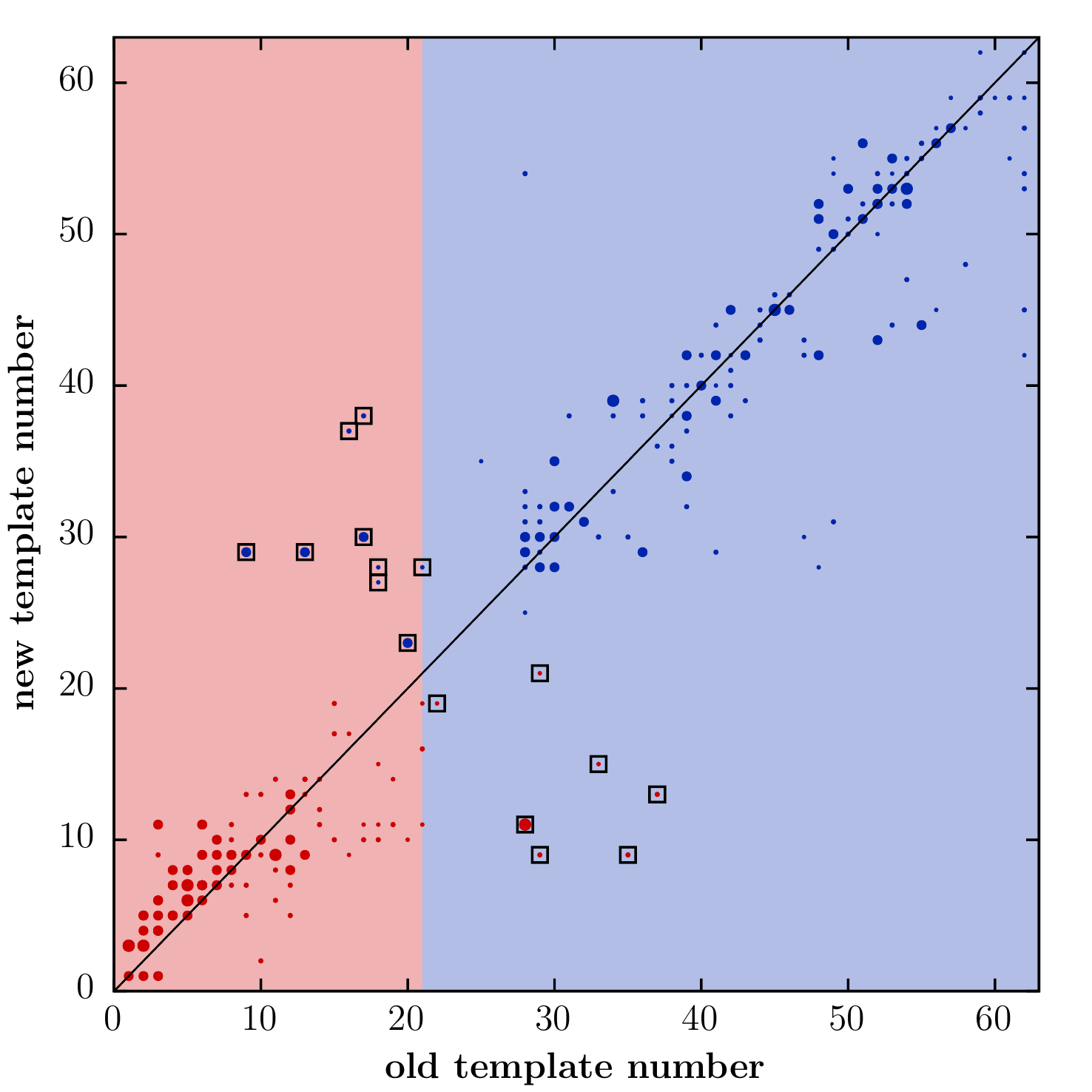}
     \caption{Comparison of best fit templates for each \textit{candidate cluster galaxy} for a rich candidate cluster at $z_{clus} = 0.35$. The blue and red
points respectively show  the new templates considered as late-type galaxies (LTGs) and early-type galaxies (ETGs). The black squares show galaxies moving from ETG to LTG and vice versa. The old template is the best fit template in the original catalogue (the redshift is free to vary during the SED fitting), while the new template is the best fit template when we force the galaxy to be at the cluster redshift ($z_{\rm phot,gal} = z_{\rm clus}$). This illustrates  that template fitting of galaxies is quite sensitive to  small changes in redshift, but that classifications are relatively stable.}
     \label{fig:templates}
   \end{figure}   
   
   \begin{figure*}[h]
\centering
   \includegraphics[width=.9\textwidth]{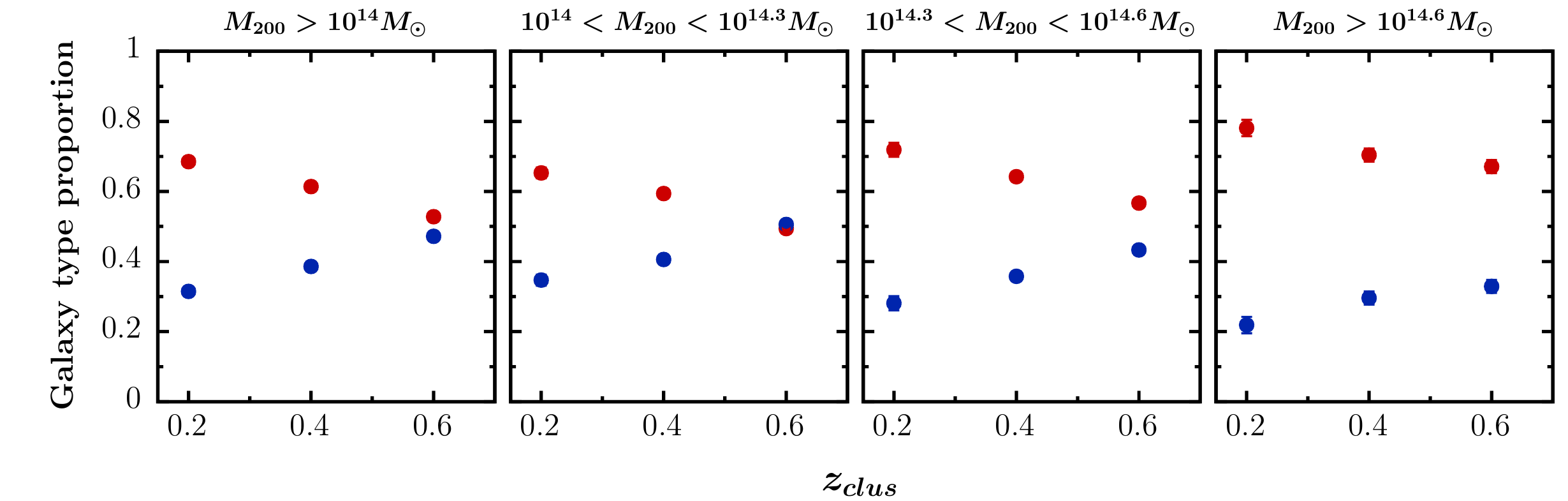}
     \caption{Mean proportion of ETGs (red) and LTGs (blue) brighter than $M^*+1.75$ in cluster candidates as a function of redshift. The left panel shows the redshift evolution for all cluster candidates with mass $M_{200} > 10^{14} \ {\rm M_\odot}$. The  other three panels segregate the cluster candidates in mass bins as defined in the text. Error bars represent the standard error on the mean.}
     \label{fig:Blue-Frac}
   \end{figure*}
We wanted to study the dependence of the GLFs with redshift and cluster mass. So we binned our cluster candidates in this 2D parameter space, and stack the clusters in each bin using the Colless method described above. As mentioned in Sect.~\ref{sec:sel_fun_Mil}, we chose to use the three redshift bins [0.1,0.3[, [0.3,0.5[, and [0.5,0.7[. These redshift bins are wide enough to be well populated and narrow enough to study the redshift dependence of the GLFs. We also use the three mass bins ]$10^{14} \ {\rm M_\odot}$, $10^{14.3} \ {\rm M_\odot}$], ]$10^{14.3} \ {\rm M_\odot}$, $10^{14.6} \ {\rm M_\odot}$], and ]$10^{14.6} \ {\rm M_\odot}$, $\infty$ [.

\subsection{Early- and late-type galaxies} \label{sect:ETG_LTG}

To better understand the properties of clusters, it is interesting to study their different galaxy populations. Ideally this problem could be dealt with in a complex manner by classifying many types of galaxies (early quiescent ellipticals, late dust-free spirals, late dusty spirals, starburst galaxies, early spirals, galaxies transiting from late spiral to early elliptical, etc.). However,  the automatic classification of galaxies in such a complicated scheme is very difficult. Indeed, the more templates/categories we try to fit the data with, the more degenerate the classification.

We used LePhare templates to classify galaxies as ETGs and LTGs. We recall that the LePhare SED fitting for cluster members is done using the ZFIX parameter with $z_{\rm gal}$ fixed at $z_{\rm clus}$. We verified that this small redshift shift does not significantly affect our ETG--LTG separation by comparing cluster best fit galaxy templates before and after fixing galaxy redshifts. This is shown in Fig.~\ref{fig:templates} for a rich cluster at $z=0.35$. We find that the best fit template is quite sensitive to a moderate change in redshift, with most galaxies having their templates changed,  yet the classification as ETG or LTG only changes for $\sim$5\% of the galaxies. This is due to the degeneracy between the spectra of ETGs and of LTGs with a dust component. It is a well-known limitation of template fitting codes, but since the proportion of these objects stays low it should not affect our conclusions.

Our classification somehow differs from the usual red-sequence (RS) classification used in most of the literature. Understanding how the two compare is important to properly compare our results with previous studies.
First, we would like to point out that previous studies also used techniques that differed from each other. Some used a simple colour cut \citep{Popesso06}. Other studies fitted a proper red sequence with a tilt in a colour-magnitude diagram, but either with a fixed slope \citep[e.g.][]{M15, DeLucia07} or varying both the slope and the intercept \citep[e.g.][] {Cerulo+16, DePropris+15}. To compare both classifications, we fitted a RS to each of our clusters in ($g - i$) versus $i$ using 
\begin{equation}
g-i = -0.0436 \times (i - m_i^*) + b,
\label{eq:RS}
\end{equation}
where the slope is fixed at -0.0436, as in \citet{M15}, and the intercept is computed at $m_i^*$, which is  the observed characteristic magnitude computed from the mean $M^*$ over our redshift range: $\langle M^* \rangle_{z} = -22.6$. As a first guess for the intercept, we interpolated the elliptical galaxy colour from \citet{Fukugita+95} to each cluster redshift and selected a wide preliminary RS with a width of 0.6 in magnitude. For the galaxies thus selected, we then fitted a RS with a free ordinate $b$ and selected galaxies at $\pm 0.3$ in magnitude around this final RS.
An example of the RS fit and galaxy selection is shown in Fig.~\ref{fig:RS_vs_Lephare} for three rich clusters at different redshifts. We checked the robustness of the fit on these clusters by changing by 0.2 the colour used for pre-selection based on \citet{Fukugita+95} and allowing the slope to vary.
        
Both checks prove our selection to be robust, as the slope only changes by 0.01 and the intercept by 0.05. Thus, only a few galaxies change their population type compared to our final RS selection.
This RS selection is compared to our ETG/LTG classification in Fig.~\ref{fig:RS_vs_Lephare}. One can see that at low redshift our selection discards part of the RS selected galaxies, especially at faint magnitudes, while the two methods agree well at high redshift. The difference at low redshift arises because  some of the galaxies in the RS are actually LTGs reddened by dust. With the observing bands we have, it is hard for LePhare to properly segregate between ETGs and dusty-LTGs for a given galaxy. However, statistically speaking, our selection is closer to a true quiescent versus  star-forming selection than the RS, because it is less polluted by these dusty-LTGs. Quantitatively, for all three clusters  99\%, 92\%, and 73\% of our ETGs are also RS galaxies. Inversely, there are, from low to high redshift  65\%, 80\%, and 96\% of RS galaxies that are classified as ETGs.

The final effect on the GLF is hard to predict since the field of a RS selection has  higher number counts than the field of an ETG selection. To properly investigate how the two methods compare with regard to the stacked GLFs, we thus built the RS GLFs for all clusters and compared the RS stacked GLFs to the ETG stacked GLFs in our nine bins of masses and redshifts. This is illustrated in Fig.~\ref{fig:stackGLF_zM-RS}. We see that at high redshift the two selections are in good agreement for all mass bins, so our conclusions concerning the evolution of the ETG GLF are not affected by the chosen selection criterion.

\section{Results} \label{GLF_results}
In this section we present the results of the GLFs of our cluster sample. We first analysed how the fraction of ETGs depends on redshift and mass. We then studied the stacked GLFs of our cluster sample. As pointed out in Sect.~\ref{sec:GLFs_cuts}, we stacked cluster candidates with a mass larger than $10^{14} \ {\rm M_\odot}$ and redshift $0.15 < z_{\rm clus} < 0.7$ in  corresponding redshift and  mass bins. We  study in particular how the parameters of our fitted Schechter functions depend on redshift and mass independently, to finally break the degeneracy between these two parameters by binning our stacked GLFs in this 2D parameter space.

\subsection{ETG fraction in clusters} \label{fraction}
\begin{figure*}[h]
\centering
    \includegraphics[width=1.0\textwidth]{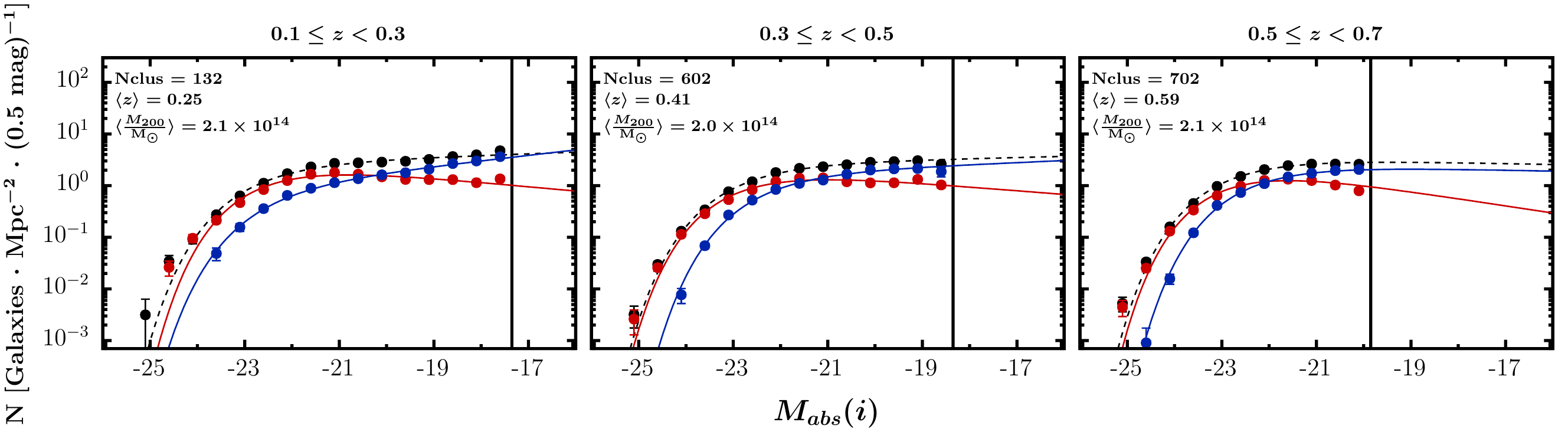}
     \caption{Redshift evolution of the i-band stacked GLF in the CFHTLS Wide fields. Black is for all galaxies, red for ETGs, and blue for LTGs. For each figure, the number of clusters, their mean redshift, and mass is indicated. The black vertical line indicates the limiting magnitude used in the fit.}
     \label{fig:stackGLF_z}
   \end{figure*}
   
   \begin{figure}[h]
   \centering
    \includegraphics[width=0.4\textwidth]{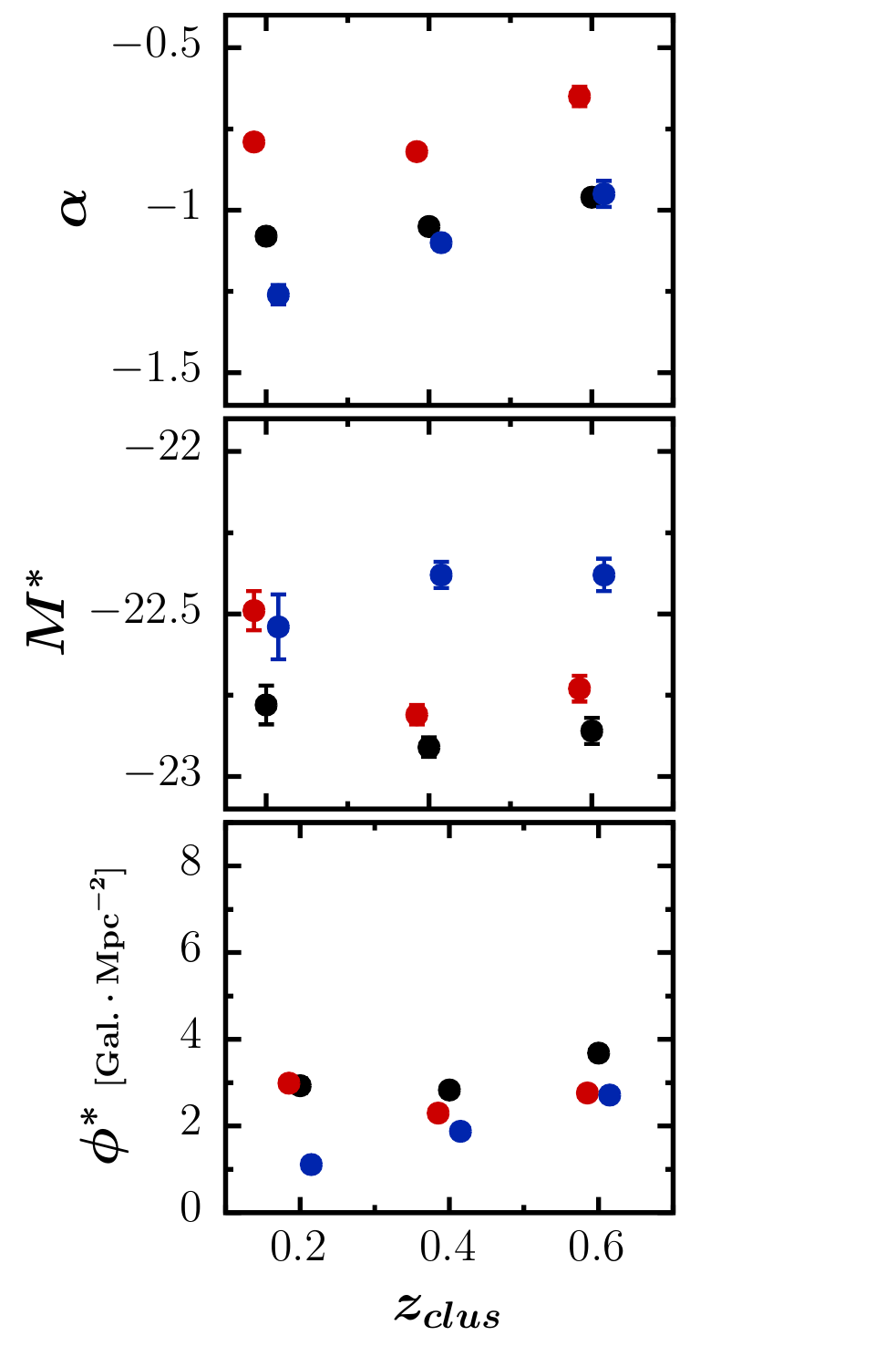}
      \caption{Evolution of the Schechter fit parameters with redshift for all galaxies (black), ETGs (red), and LTGs (blue).}
     \label{fig:schech_z}
   \end{figure}

Using  the methods presented in Sects.~\ref{GLF} and \ref{sect:ETG_LTG} we are able, for each individual cluster candidate, to compute the number of ETGs and LTGs down to $M^*+1.75$. We can thus compute the fraction of ETGs in each candidate cluster. As for the richness computation, counting galaxies down to $M^*+1.75$ enables us to cover the redshift range $0.15 < z_{\rm clus} < 0.7$ homogeneously, as all clusters are complete at this magnitude limit in this redshift range. 

To study the dependence on redshift and mass, we first bin our cluster candidates with mass larger than $10^{14} \ {\rm M_\odot}$ in three mass bins. For each mass bin, cluster candidates are binned in redshift and the mean ETG fraction and LTG fraction is computed for clusters. The results are shown in Fig.~\ref{fig:Blue-Frac}, where the galaxy type proportions are compared to the redshift dependence when no segregation in mass is performed. The error bars represent the standard error on the mean.

This analysis shows a clear increase in the fraction of ETGs in cluster candidates ($M_{200} > 10^{14} \ {\rm M_\odot}$) with redshift, from $53 \pm 1 \%$ at $z=0.6$ to $69 \pm 1 \%$ at $z=0.2$.

When segregating our cluster candidates into mass bins, we find that the fraction of ETGs at a given redshift is strongly dependent on mass. At $z=0.2$, the mean proportion of ETGs in our cluster candidates is $66 \pm 2 \%$ for the lowest mass bin ($14 < \log{M_{200}/{\rm M_\odot}} \le 14.3$), $72 \pm 2 \%$ for the intermediate mass bin ($14.3 < \log{M_{200}/{\rm M_\odot}} \le14.6$), and $78 \pm 2 \%$ for the highest mass bin ($\log{M_{200}/{\rm M_\odot}} > 14.6$). At $z=0.6$, the mean proportion of ETGs in our cluster candidates is $49 \pm 1 \%$ for the lowest mass bin, $57 \pm 1 \%$ for the intermediate mass bin, and $67 \pm 2 \%$ for the highest mass bin. 

\subsection{GLFs of clusters stacked in redshift} \label{GLF:stackresz}
   
Computing the stacked GLFs of our cluster candidates in bins of redshift, we are able to study the redshift evolution of our Schechter fit parameters. As mentioned in Sect.~\ref{GLF:stackmethod}, the Schechter fits of individual cluster candidates are too noisy to study their redshift or mass dependence. We thus stack our cluster candidates using the Colless method. 

The results are shown in Fig.~\ref{fig:stackGLF_z}, where for each panel we indicate the number of cluster candidates stacked, their mean redshift, and mean mass. The best fit parameters are listed in Table \ref{tab:schech_z}, where we also provide the absolute magnitude limit to which the fit is performed ($compl$) and a goodness of fit parameter ($p$) defined in Eq.~\ref{eq:p}. 

In our three redshift bins, the Schechter fits to the GLFs of all the galaxies, ETGs and LTGs all converged with a goodness of fit parameter $p > 0.94$. We study the evolution of the three fitted parameters: the normalisation $\phi ^*$, the characteristic absolute magnitude of the knee $M^*$, and the faint-end parameter $\alpha$. Figure~\ref{fig:schech_z} summarizes the results, where each parameter is plotted against redshift. Figure~\ref{fig:Mstar_vs_alpha-z} in Appendix~\ref{sect:alpha-vs_Mstar} shows the confidence ellipses on the values of $M^*$ and $\alpha$.

For the faint-end parameter $\alpha$, there is a mild flattening with decreasing redshift for  the ETG and the LTG populations. The ETG population GLF faint end flattens from $\alpha_{\rm ETG} = -0.65 \pm 0.03$ at $z=0.6$ to $\alpha_{\rm ETG} = -0.79 \pm 0.02$ at $z=0.2$, i.e. a difference of 0.14 with a significance of $3.9 \sigma$. The LTG population GLF faint end steepens from $\alpha_{\rm LTG} = -0.95 \pm 0.04$ at $z=0.6$ to $\alpha_{\rm LTG} = -1.26 \pm 0.03$ at $z=0.2$, i.e. a difference of 0.31 with a significance of $6.2 \sigma$. The overall population GLF faint end steepens from $\alpha_{\rm all} = -0.96 \pm 0.02$ at $z=0.6$ to $\alpha_{\rm all} = -1.08 \pm 0.02$ at $z=0.2$, i.e. a difference of 0.12 with a significance of $4.2 \sigma$.

The absolute magnitude characteristic parameter $M^*$ is compatible with no evolution for all galaxies and LTGs. The ETG population $M^*$ redshift dependence is compatible with passive evolution. The normalization $\phi ^*$ of the overall population  decreases with decreasing redshift, with a significance of $2.8\sigma$. This is due to the LTG population, which follows a similar evolution (with significance $> 10 \sigma$), while the ETG population shows no redshift dependence.
 
\subsection{GLFs of clusters stacked in mass bins} \label{GLF:stackresM}

\begin{figure*}
\centering
    \includegraphics[width=1.0\textwidth]{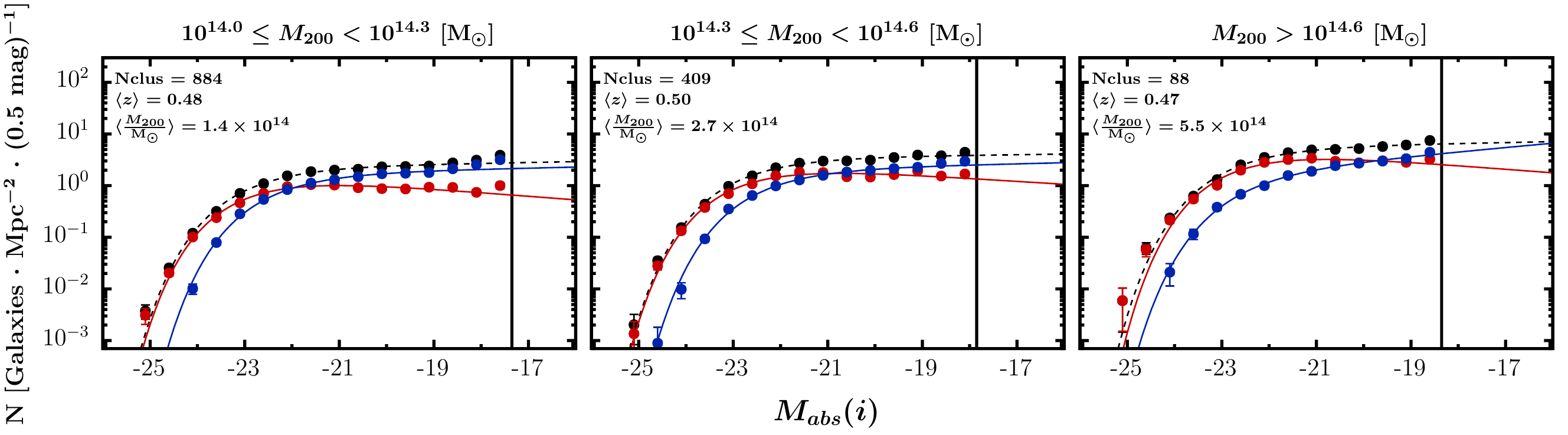}
    \caption{Mass dependence of the i-band stacked GLF in the CFHTLS Wide fields. Black is for all galaxies, red for ETGs, and blue for LTGs. For each figure the number of clusters, their mean redshift, and mass is indicated. The black vertical line indicates the limiting magnitude used in the fit.}
     \label{fig:stackGLF_M}
   \end{figure*}

   \begin{figure}[h]
   \centering
    \includegraphics[width=0.4\textwidth]{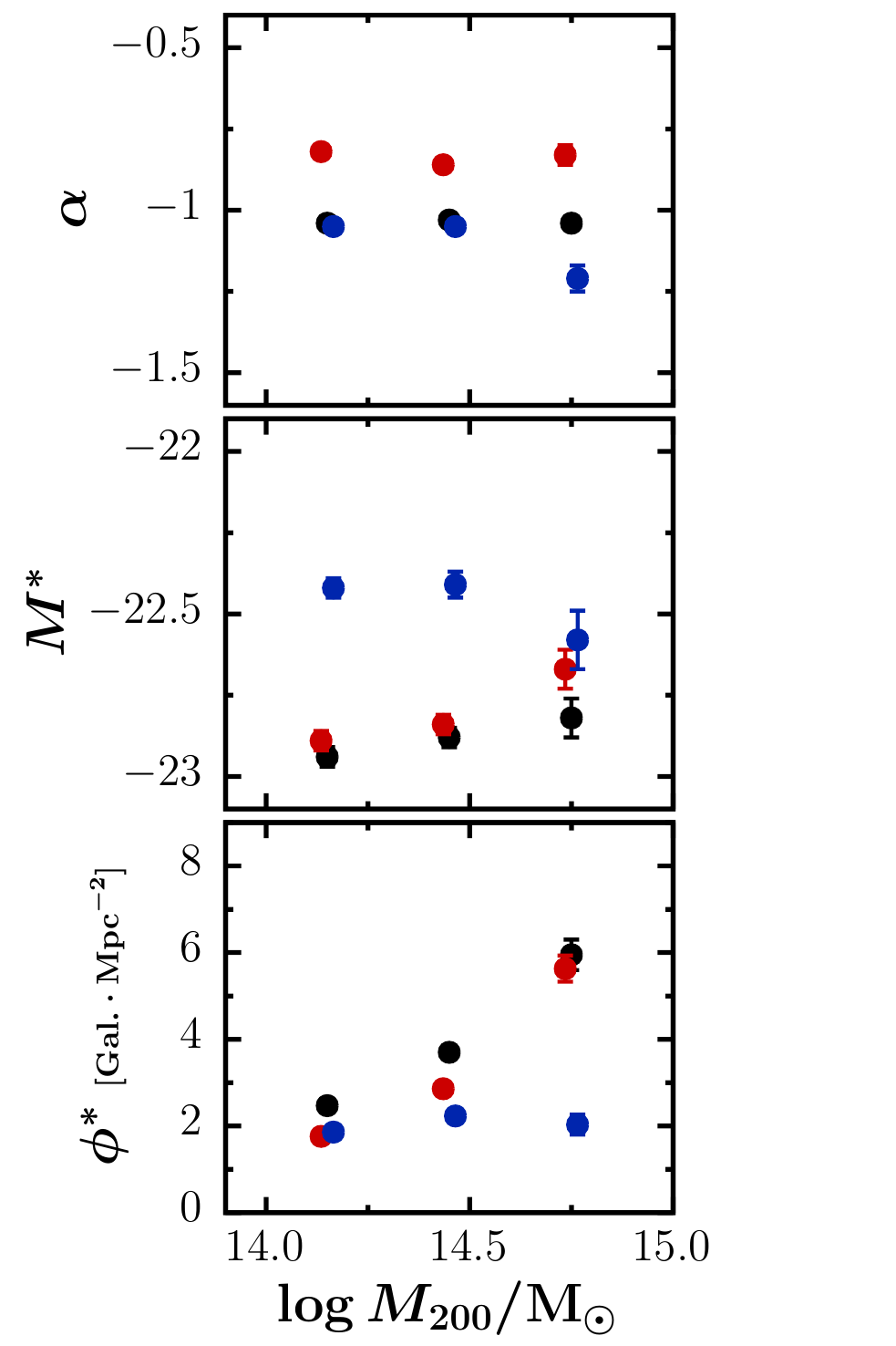}
         \caption{Dependence of the Schechter fit parameters on mass for all galaxies (black), ETGs (red), and LTGs (blue).}
     \label{fig:schech_M}
   \end{figure}
   
We also make use of our mass inference derived from the mass-richness calibration described in Sect.~\ref{sec:richness} to study the dependence of our fitted parameters with the mass of the cluster candidates. We use the same stacking method as in the previous section. 

The results are shown in Fig.~\ref{fig:stackGLF_M}, where  for each panel  we indicate the number of cluster candidates stacked, their mean redshift, and mean mass. The best fit parameters are listed in Table \ref{tab:schech_M}, where we also provide the absolute magnitude limit to which the fit is performed ($compl$)  and the goodness of fit parameter $p$  defined in Eq.~\ref{eq:p}.\\
\indent In our three mass bins, the Schechter fits to the GLFs of all the galaxies, of ETGs, and LTGs all converged with a goodness of fit parameter $p > 0.72$. Using these fits, we study the evolution of the Schechter parameters. Figure~\ref{fig:schech_M} summarizes the results with each parameter plotted against mass. Figure~\ref{fig:Mstar_vs_alpha-M} in Appendix~\ref{sect:alpha-vs_Mstar} shows the confidence ellipses on the values of $M^*$ and $\alpha$.

For the faint-end parameter $\alpha$, we observe a steepening with increasing mass for the LTG population, from $\alpha_{\rm LTG} = -1.05 \pm 0.02$ for $14 < \log{(M_{200})} \le 14.3$ to $\alpha_{\rm LTG} = -1.21 \pm 0.04$ for $\log{(M_{200})} > 14.6$, with a significance of $3.6 \sigma$. The ETG population and the full galaxy population show no clear evidence for a mass dependence of the faint end. The absolute magnitude characteristic parameter $M^*$ is compatible with no mass dependence for the full galaxy population or  for the LTGs. For the ETG population, $M^*$ decreases with increasing mass with a significance of $\sim 3 \sigma$.

The normalization $\phi ^*$ depends on mass for the overall galaxy population, with a significance of $9.7 \sigma$. This is expected since the normalization of the GLFs is directly related to richness, and richness is a proxy for mass. This is also true for the ETG population, which shows a clear dependence (with a $>12 \sigma$ significance) meaning that there are more ETGs in massive clusters. On the other hand, the normalization of the GLF of the LTG population seems independent of mass.

\subsection{Breaking the degeneracy: GLFs of clusters stacked in mass-redshift bins} \label{sec:break_degen}

When studying a sample of clusters, there is usually a degeneracy between mass and redshift. Indeed, because more massive clusters are rarer, by binning our clusters in redshift space only we actually sample a population that is more representative of low-mass clusters. However, low-mass clusters are more difficult to detect at high redshift, so the two effects act oppositely and it is difficult to know which one dominates.

To break this degeneracy we need to bin in the 2D parameter space of both mass and redshift. To date this had not been possible because studies of the evolution of cluster GLFs with redshift have been limited to small samples, and extensive studies of the mass dependence concern only low-redshift objects \citep[see e.g.][on SDSS data]{Lan16}.

 \begin{figure*}
         \centering 
          \includegraphics[width=1.0\textwidth]{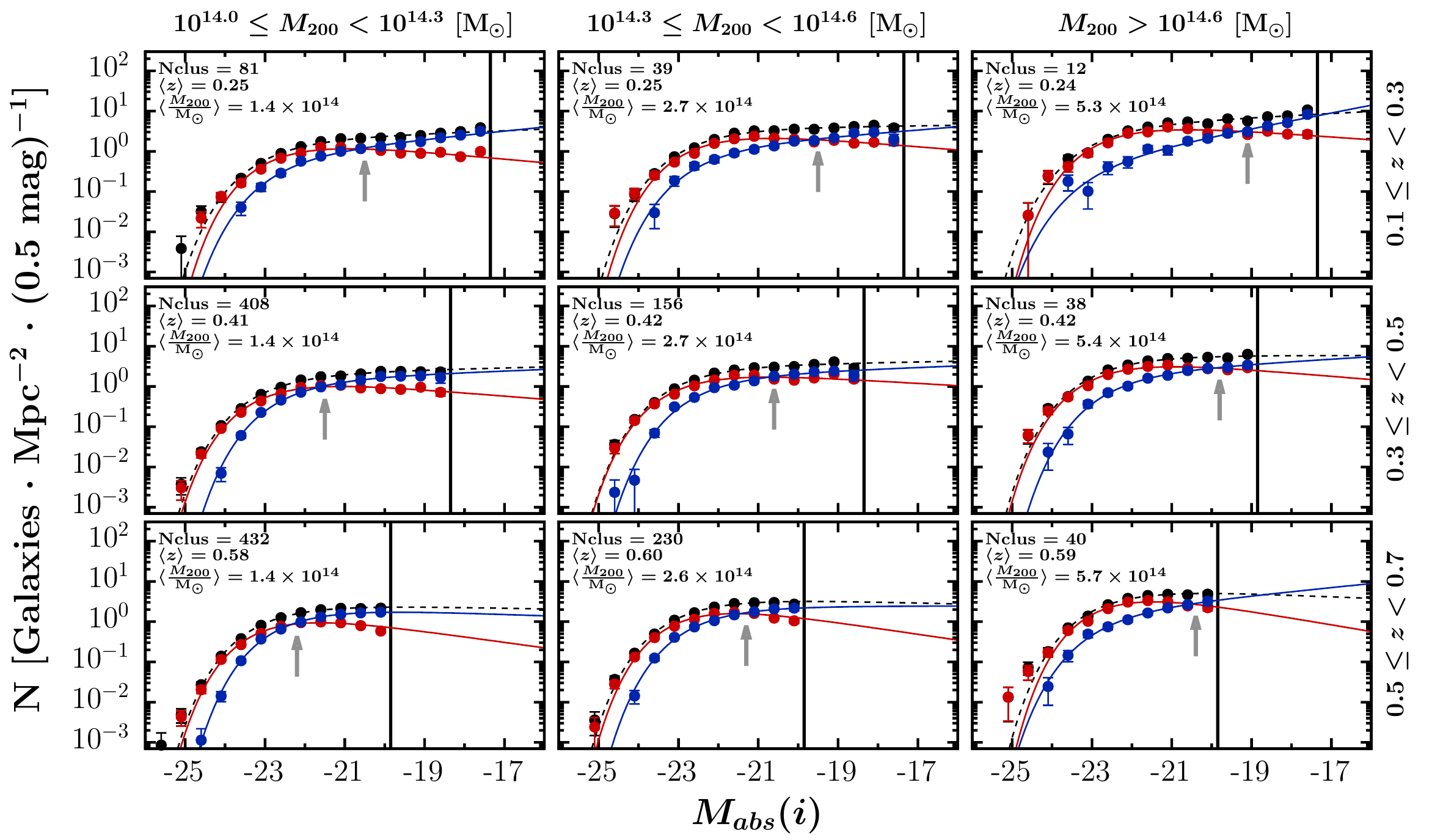}
     \caption{Redshift and mass co-evolution of the i-band stacked GLFs in the W1 field of the CFHTLS in bins of mass and redshift. Black is for all galaxies, red for ETGs, and blue for LTGs. Indicated in  each panel are the mean redshift and mass, and   the number of cluster candidates in the bin. The black vertical line indicates the limiting magnitude used in the fit. The grey arrows indicate the absolute magnitudes where the GLFs of the ETGs and LTGs intersect.}
     \label{fig:stackGLF_zM3}
 \end{figure*}

\begin{figure*}
\centering
   \includegraphics[width=.8\textwidth]{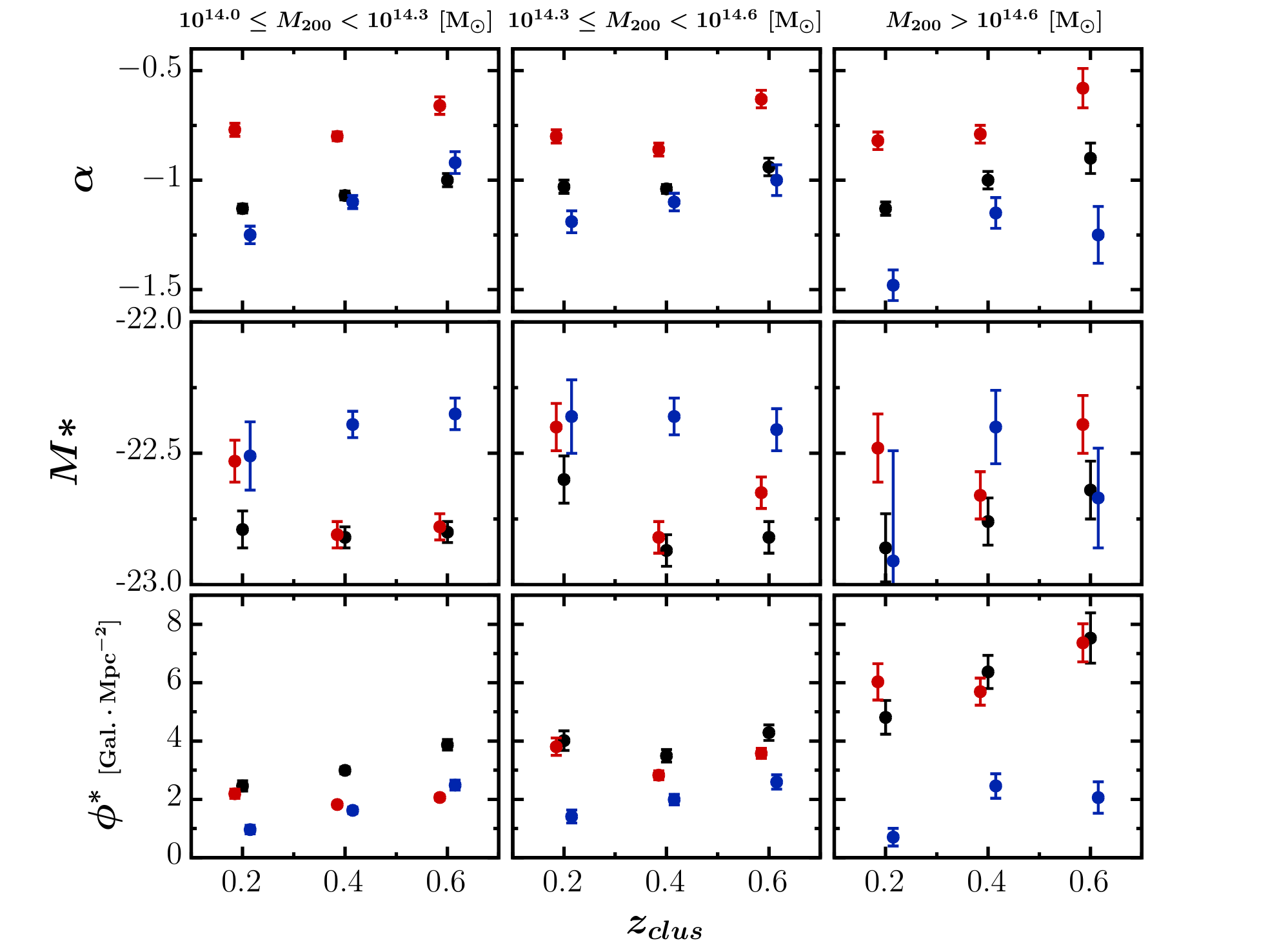}
     \caption{Evolution of the Schechter fit parameters with redshift and mass for all galaxies (black), ETGs (red), and LTGs (blue).}
     \label{fig:fitparam}
   \end{figure*}

The large size of our cluster candidate catalogue and the depth of the CFHTLS enable us to carry out such an analysis. Indeed, considering the four Wide fields of the CFHTLS, we detect 1371 cluster candidates with $M_{200} > 10^{14} \ {\rm M_\odot}$, $S/N > 4$ and $z < 0.7$. Since we computed a robust selection function from the simulations, we know how representative of the true underlying cluster population our sample is in terms of completeness. Our catalogue is $>90\%$ pure in the redshift range considered and at worse $\sim 50\%$ complete (for the lowest mass and highest redshift bin). Overall, the 2D bins are  well populated, with 12 cluster candidates in the less populated bin and 160 clusters per bin in the mean. 

The GLFs obtained in the three redshift bins and three mass bins are shown in Fig.~\ref{fig:stackGLF_zM3}, where for each panel  we indicate the number of cluster candidates stacked, their mean redshift, and mean mass. The best Schechter fit to each galaxy population is overplotted. A grey arrow indicates the absolute magnitude where the red and blue GLFs intersect. We can see that for all cluster masses the intersection moves towards brighter magnitudes as the redshift increases. Therefore the proportion of faint LTGs increases with decreasing redshift. For a given redshift, the grey arrow moves towards fainter magnitudes as the cluster mass increases, implying that in massive clusters the contribution of bright ETGs is higher than in low-mass clusters. At fixed mass, the slope flattens ($\alpha$ more negative) for ETGs and steepens ($\alpha$ more negative) for LTGs from high to low redshift. This leads to the magnitude at which the two populations cross getting brighter at higher redshift. We see the opposite evolution of the slopes with increasing mass compared to the redshift evolution, so that the crossing happens at brighter magnitudes for low-mass clusters.

The best fit parameters are presented in Table \ref{tab:schech_zM}, where we also provide the absolute magnitude limit $compl$ to which the fit is performed and the goodness of fit parameter $p$ defined in Eq.~\ref{eq:p}.

The Schechter fits to the GLFs of the full galaxy population and ETG population are obtained with goodness of fit parameters $p > 0.94$. For the LTG population, the  goodness of fit parameter is lower on average, but always greater than 0.68. Using these fits, we can study the evolution of the fitted parameters. Figure~\ref{fig:fitparam} summarizes the results, by plotting $\alpha$, $M^*$, and $\phi^*$ as a function of redshift for each of the three mass bins. Figure~\ref{fig:Mstar_vs_alpha-M} in Appendix~\ref{sect:alpha-vs_Mstar} shows the confidence ellipses on the values of $M^*$ and $\alpha$.

For ETGs, the faint-end  slope $\alpha$ slightly flattens with decreasing redshift in the three mass bins. In the lowest mass bin ($14 < \log{(M_{200})} \le 14.3$), it decreases from $\alpha_{\rm ETG} = -0.66 \pm 0.04$ at $z=0.6$ to $\alpha_{\rm ETG} = -0.77 \pm 0.03$ at $z=0.2$, i.e. a difference of 0.11 with a significance of only $2.2 \sigma$. The same trend is observed for the intermediate mass bin ($14.3 < \log{(M_{200})} \le 14.6$), the faint-end slope flattening from $\alpha_{\rm ETG} = -0.63 \pm 0.04$ at $z=0.6$ to $\alpha_{\rm ETG} = -0.80 \pm 0.03$ at $z=0.2$, i.e. a difference of 0.17 with a significance of $3.4 \sigma$, and in the highest mass bin ($\log{(M_{200})} > 14.6$) with a flattening from $\alpha_{\rm ETG} = -0.58 \pm 0.09$ at $z=0.6$ to $\alpha_{\rm ETG} = -0.82 \pm 0.04$ at $z=0.2$, i.e. a difference of 0.24 with a significance of $2.4 \sigma$. These results also hint to a mass dependence of the faint-end slope at a given redshift. Indeed, in the high-redshift bin ($z=0.6$) the slope is flatter for the low-mass clusters, with a difference of 0.08. The opposite is observed in the low-redshift bin ($z=0.2$) where the slope is flatter for the high-mass clusters, with a difference of 0.05.  However, these hints are detected at a low significance level and would thus need extended mass and/or redshift coverage to be verified.

For LTGs the faint-end slope decreases with decreasing redshift in the two lowest mass bins and shows no clear evolution in the highest mass bin.
In the lowest mass bin, it steepens from $\alpha_{\rm LTG} = -0.92 \pm 0.05$ at $z=0.6$ to $\alpha_{\rm LTG} = -1.25 \pm 0.04$ at $z=0.2$, i.e. a difference of 0.33 with a significance of $5.2 \sigma$. The same trend is observed for the intermediate mass bin, where the faint-end slope steepens from $\alpha_{\rm LTG} = -1.00 \pm 0.07$ at $z=0.6$ to $\alpha_{\rm LTG} = -1.19 \pm 0.05$ at $z=0.2$, i.e. a difference of 0.19 with a significance of $2.2 \sigma$. 

The faint-end slope of the overall population also steepens with decreasing redshift in the three mass bins. In the lowest mass bin, it steepens from $\alpha_{\rm all} = -0.95 \pm 0.03$ at $z=0.6$ to $\alpha_{\rm all} = -1.09 \pm 0.02$ at $z=0.2$, i.e. a difference of 0.14 with a significance of $3.9 \sigma$. The same trend is observed for the intermediate mass bin ($14.3 < \log{(M_{200})} \le 14.6$) but with a lower significance level, the faint-end slope steepening from $\alpha_{\rm all} = -0.94 \pm 0.04$ at $z=0.6$ to $\alpha_{\rm all} = -1.03 \pm 0.03$ at $z=0.2$, i.e. a difference of 0.09 with a significance of $1.8 \sigma$, and in the highest mass bin ($\log{(M_{200})} > 14.6$) with a steepening from $\alpha_{\rm all} = -0.90 \pm 0.07$ at $z=0.6$ to $\alpha_{\rm all} = -1.13 \pm 0.03$ at $z=0.2$, i.e. a difference of 0.23 with a significance of $3 \sigma$.

In Fig.~\ref{fig:fitparam}, we see an evolution of $M^*$ compatible with both passive evolution and no evolution for ETGs in the two lowest mass bins. No particular evolution is seen for other populations.

When looking at the normalization parameter $\phi^*$, we observe a decrease with decreasing redshift for the LTG population, respectively at $6.8 \sigma$, $3.5\sigma$, and $2.2\sigma$ in the low-, intermediate-, and high-mass bins, between $z=0.6$ and $z=0.2$. The ETG population shows a rather constant normalisation with redshift, so that the overall population normalisation tends to follow the LTG evolution.

These results are discussed in Sect.~\ref{sec:discussion}.

\section{Discussion}\label{sec:discussion}

\subsection{ETG fraction}
We first consider the evolution of cluster galaxy types with redshift for galaxies brighter than $M^* + 1.75$. The fraction of bright ETGs increases with decreasing redshift whatever the mass. The evolution is stronger for the low-mass clusters, with a change of about 20\% in the fraction of ETGs over the probed redshift range, against only 10\% for high-mass clusters.  At fixed redshift, however, we see a higher fraction of ETGs in high-mass clusters compared to low-mass clusters. These two pieces of information can be interpreted as a different behaviour between the high- and low-mass clusters. The low-mass clusters undergo a significant evolution of the bright ETG fraction over the redshift range $0.1<z<0.7$; however, this ETG fraction is  lower than that of high-mass clusters, which probably underwent a similar evolution earlier in their history, at $z>0.7$. In this scenario, the massive clusters are more evolved than the low-mass clusters, which  are still evolving towards higher masses either by accreting ETGs from preprocessed groups or by an evolution of LTGs into ETGs. 

\subsection{GLF redshift evolution}
We focus on the variations of the Schechter parameters with cluster redshift. To study the GLF behaviour, we differentiate between galaxies with $M_{\rm abs}(i) < M^* + 1.75$, which we call bright galaxies, and galaxies with $M_{\rm abs}(i) > M^* + 1.75$, which we call faint galaxies. 

\begin{figure}[!]
\centering
   \includegraphics[width=0.5\textwidth]{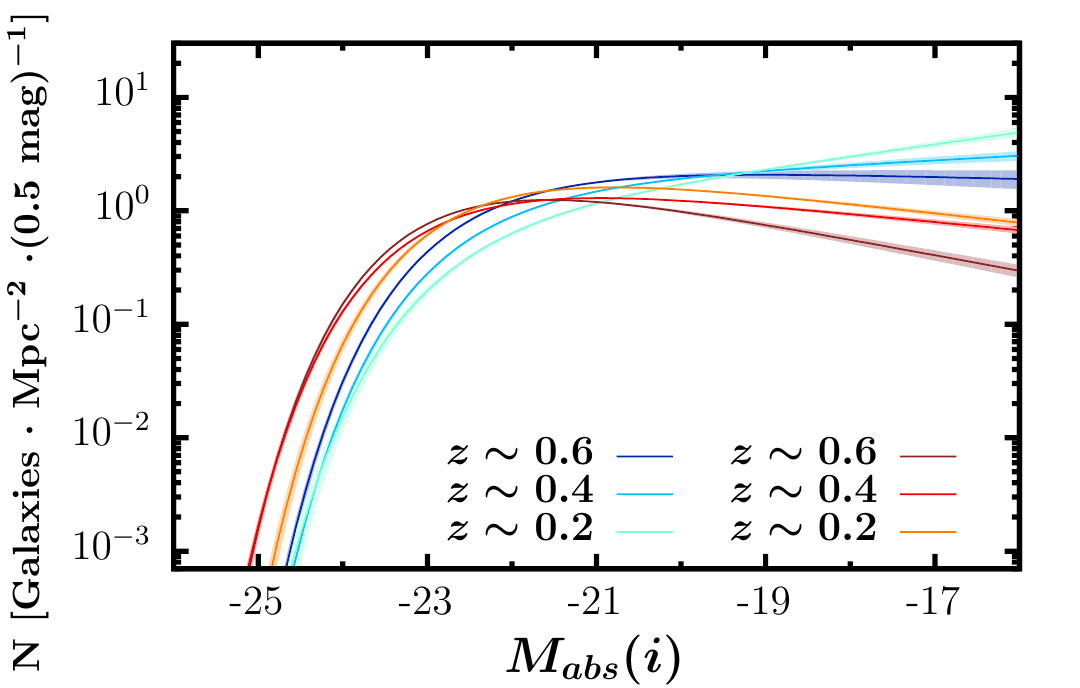}
     \caption{Redshift evolution of the Schechter fit of ETG (orange, red, and brown) and LTG (green, light blue, and deep blue) GLFs. The shaded areas are the 68\% confidence interval on the fit.}
     \label{fig:GLF-comp-z}
   \end{figure}

In Sect.~\ref{GLF:stackresz} we see that when clusters of all masses are included, $\alpha$ steepens and $\phi^*$ decreases with decreasing redshift for the LTG population, while $M^*$ remains constant. This implies that there are more faint LTGs ($\alpha_{\rm LTG}$ more negative) and less bright LTGs (smaller $\phi^*$) at $z=0.2$ than at $z=0.6$. This can be seen in Fig.~\ref{fig:GLF-comp-z}, which shows how the Schechter function changes with redshift for both ETGs and LTGs. This is also in agreement with the evolution of galaxy types with redshift. For the ETG population, we observe a slight flattening of $\alpha_{\rm ETG}$ and an increase in $M^*$ between $z=0.6$ and $z=0.2$ (compatible with a passive evolution) and a constant $\phi^*$. These effects combine so that there are more faint ETGs ($\alpha_{\rm ETG}$ flattens) at $z=0.2$ than at $z=0.6$, while the number of bright ETGs is slightly higher at $z=0.6$. The passive evolution is consistent with a single stellar population model of \citet{BC03}, with a single burst of star formation at $z=3$. The redshift evolution of the faint-end slope parameter $\alpha_{\rm ETG}$ is in agreement with previous studies \citep[e.g.][and references therein]{Martinet+17}. \citet{Zhang+17} recently found a compatible redshift dependence of the faint-end slope for the red-sequence GLF using Dark Energy Survey (DES) data. We also note that when comparing our results to those of  \citet{Martinet+17}, our faint-end slope parameter $\alpha$ is slightly steeper at low redshift than theirs, but within the error bars. 
However,  some studies--most recently \citet{DePropris+15} and \citet{Cerulo+16}--see no clear redshift evolution in the faint-end slope of the red-sequence GLF. Both studies build the stacked GLFs of high-redshift clusters and find faint-end slope values compatible with low redshift values. 

\citet{Cerulo+16} has studied a sample of nine very high-redshift clusters ($0.8<z<1.5$) and find a somewhat weaker evolution of the faint end than ours. Namely, they measured $\alpha_{\rm RS}=-0.889\substack{+0.009 \\ -0.001}$ for red-sequence galaxies. The difference with our study could be explained by the low number of clusters they consider, together with the different range of redshift, or by a contamination of our cluster candidate sample by false detections. The latter option is  unlikely, however,  as we ensured a 90\% purity in our selected sample.
\citet{DePropris+15} studied a sample of four clusters at a mean redshift of $z = 1.25$. When looking at the stacked red-sequence GLFs, they seem to find a weaker evolution than ours. Even though no error bars on the faint-end parameter is given, when looking at their M* versus $\alpha$ confidence ellipses, their best fit seems to be compatible with $\alpha_{\rm RS} \sim -0.5$, so there should be less than a $1 \sigma$ discrepancy with our results.

Two different scenarios could explain these observations. In the first, the evolution of cluster GLFs with redshift is driven by stripping of LTGs (explaining the decrease in bright LTG density and the increase in faint LTG density) and accretion of preprocessed galaxy groups (explaining the increase in faint ETG density). In the second, the evolution results from the combination of accretion of faint LTGs from the field and quenching of bright LTGs into ETGs of slightly fainter magnitudes. This second scenario also agrees with the model of \citet{Peng+10}.
   
\subsection{GLF mass dependence}

\begin{figure}[!]
\centering
   \includegraphics[width=0.5\textwidth]{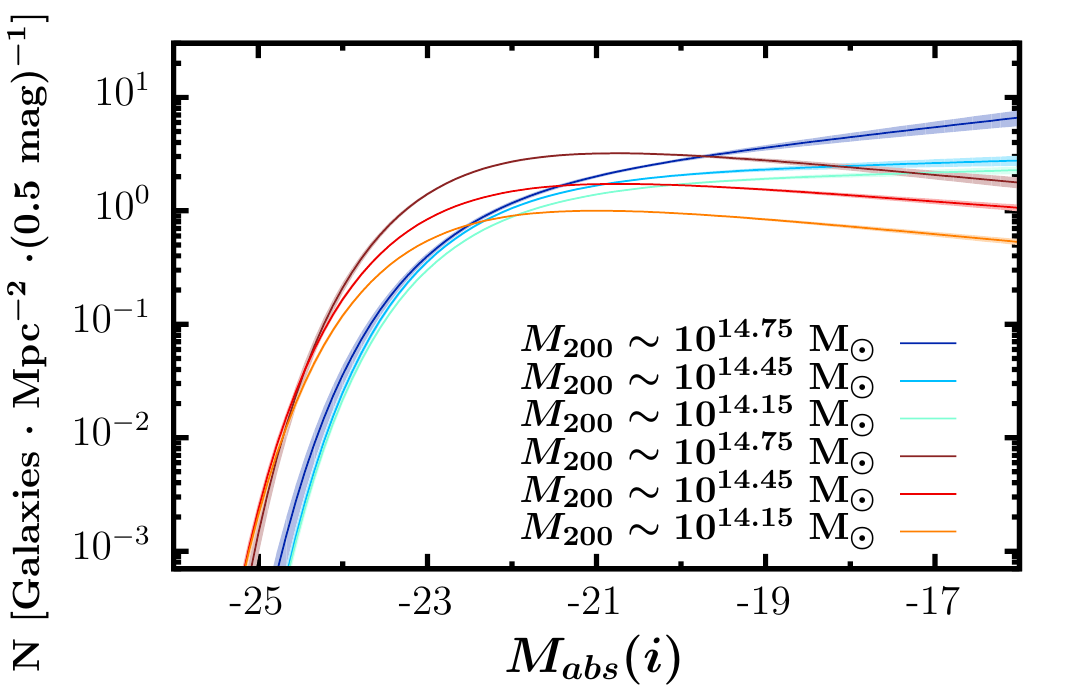}
     \caption{Mass dependence of the Schechter fit of ETG (orange, red, and brown) and LTG (green, light blue, and deep blue) GLFs. The shaded areas are the 68\% confidence interval on the fit.}
     \label{fig:GLF-comp-M}
   \end{figure}

We see in Sect.~\ref{GLF:stackresM} that when clusters are binned in mass to compute the stacked GLFs, $\alpha$ is mass dependent for LTGs only, while $M^*$ shows a slight mass dependence for ETGs alone (Fig.~\ref{fig:GLF-comp-M}). This is mostly in agreement with the \citet{Lan16} results on SDSS data where no mass dependence was found for $\alpha$ or $M^*$ for both galaxy populations in the mass range studied here. \citet{Zhang+17} also recently reported no evidence for a mass dependence of both parameters on red-sequence galaxies. Moreover, the \citet{Lan16} value of $\alpha_{\rm LTG}$ is somewhat steeper than ours, being around $-1.5$, whatever the cluster mass. This suggests that the evolution of faint LTGs we see in our study continues toward lower redshifts. The normalisation $\phi^*$ is strongly mass dependent for ETGs, with more massive clusters having a higher normalization, while it is independent of mass for LTGs. This shows that the number of bright ETGs is a good proxy of the mass, contrary to the LTGs. The combination of the $\alpha$ and $\phi^*$ mass dependence implies that more massive clusters contain more faint LTGs than low-mass clusters do, and also contain more bright ETGs. 

This suggests that accretion of faint LTGs from the field could be more efficient in more massive clusters. The constant number of bright LTGs with cluster mass disfavours stripping or disruption of bright LTGs into faint LTGs. Indeed, these effects are driven by the environment density and are thus expected to be mass dependent.

\subsection{GLF redshift evolution with redshift and mass}
   \begin{figure*}
\centering
   \includegraphics[width=1.0\textwidth]{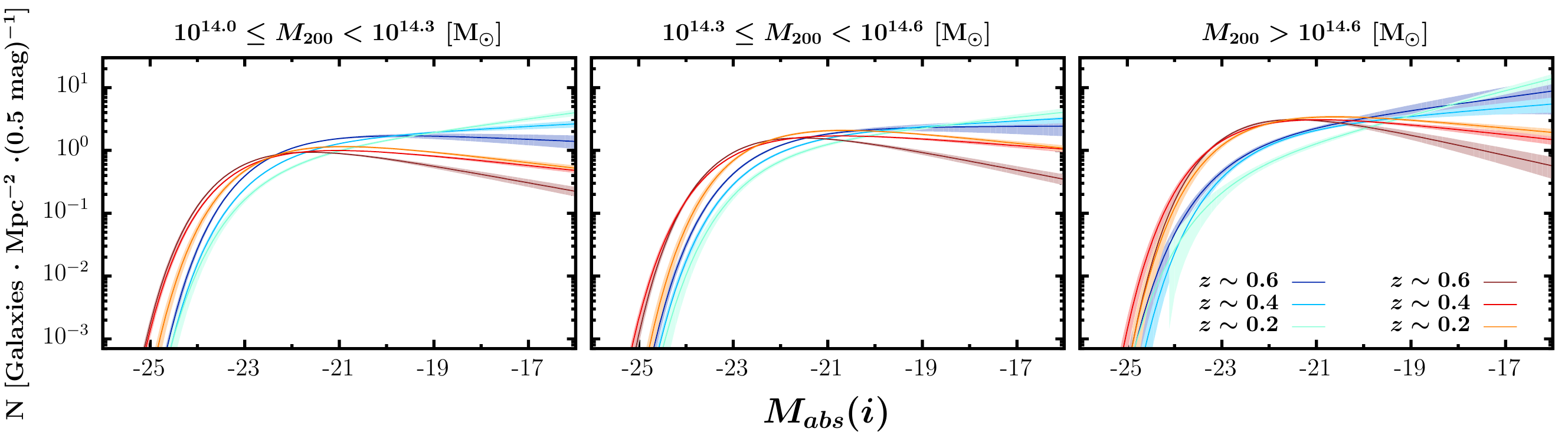}
     \caption{Redshift evolution of the Schechter fit of ETG (orange, red, and brown) and LTG (green, light blue, and deep blue) GLFs at fixed mass. The three panels are for the three different mass bins. The shaded areas are the 68\% confidence interval on the fit.}
     \label{fig:GLF-comp-zM}
   \end{figure*}
   
    \begin{figure*}
\centering
   \includegraphics[width=1.0\textwidth]{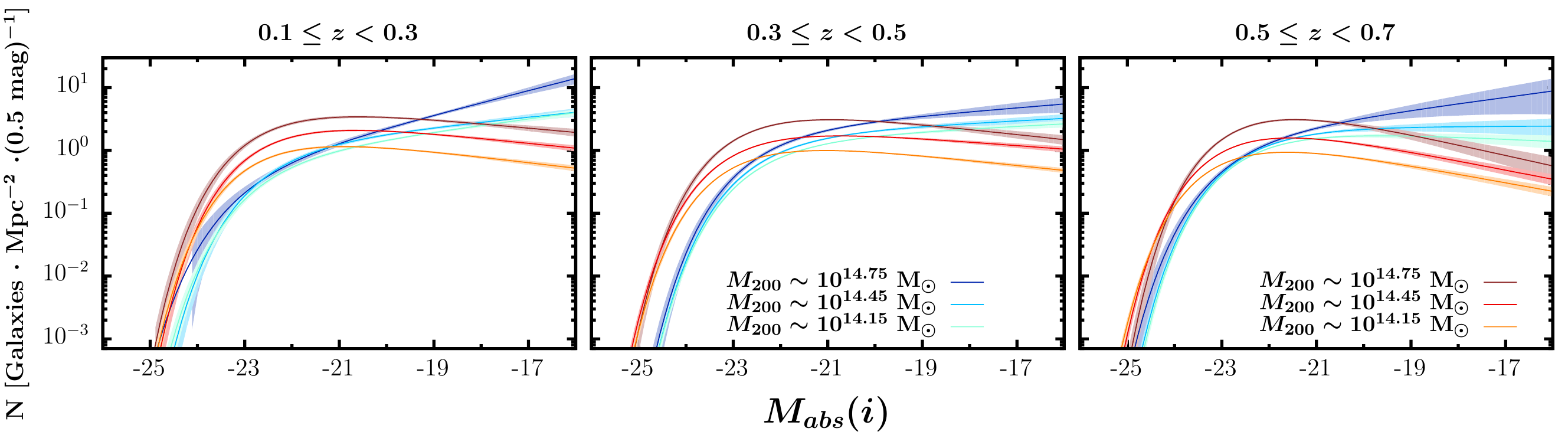}
     \caption{Mass dependence of the fit of ETG (orange, red, and brown) and LTG (green, light blue, and deep blue) GLFs at fixed redshift. The three panels are for the three different redshift bins. The shaded areas are the 68\% confidence interval on the fit.}
     \label{fig:GLF-comp-zM2}
   \end{figure*}  
As already stated, the dependence of GLFs with redshift or mass, although it gives insights into the physical processes happening in clusters, suffers from the degeneracy between these two parameters. By breaking the degeneracy, we can look at how the GLFs depend on redshift for given cluster masses, or inversely look at how they depend on mass for given cluster redshifts.

When looking at the redshift evolution for different mass bins, we have seen in Sect.~\ref{sec:break_degen} that the faint-end slope $\alpha_{\rm ETG}$ of ETG GLFs slightly flattens with decreasing redshift for all cluster masses, but the more massive the cluster, the more it flattens between $z=0.6$ and $z=0.2$ (see Fig.~\ref{fig:fitparam}). This is even clearer when directly comparing the Schechter function evolution at different masses (see Fig.~\ref{fig:GLF-comp-zM}). The combined evolution of the three parameters implies that there is a significant increase in the number density of faint ETGs with decreasing redshift, and that it is stronger for higher masses. This means that more massive clusters see their faint ETG population grow faster between $z=0.6$ and $z=0.2$ than less massive ones. For LTGs, $\alpha_{\rm LTG}$ flattens with decreasing redshift for the lowest mass bin, while it shows no clear redshift evolution in the two higher mass bins. At the same time, the normalization $\phi^*$ decreases with decreasing redshift, this effect also being stronger for lower mass clusters. The number density of bright LTGs decreases with decreasing redshift for all cluster masses.

When looking at the mass dependence for different redshift bins (see Fig.~\ref{fig:GLF-comp-zM2}), we see that the shapes of the ETG GLFs at a given redshift are independent of mass. Only the normalisation changes: more massive clusters have more ETGs, both bright and faint. For LTGs $\alpha_{\rm LTG}$ gets steeper as the mass increases whatever the redshift. Also, high-mass clusters have more faint LTGs in our redshift range. We see that the number density of bright LTGs does not depend on mass, whatever the redshift. We find more faint LTGs in high-mass clusters, but a larger increase with decreasing redshift for the low mass, which suggests that high-mass clusters have already accreted most of their environment at higher redshift. We note that this last conclusion becomes obvious only when breaking the mass-redshift degeneracy.

\subsection{Physical interpretation}

To sum up, the number of faint ETGs increases with decreasing redshift, the effect being stronger at higher masses. The number of faint LTGs tends to be higher in higher mass clusters whatever the redshift, but only low-mass clusters see their number density of faint LTGs increase in the redshift range studied. The number density of bright LTGs decreases with decreasing redshift, whatever the mass. 

This favours the scenario where the red-sequence formed at $z>0.7$ with low to no evolution of bright ETGs at $z<0.7$. This red-sequence is then enriched between $z=0.7$ and $z=0.15$ through quenching of bright LTGs into ETGs of slightly fainter magnitudes. This quenching is more efficient in high-mass clusters. At the same time, accretion of faint LTGs is more efficient at low mass, as high-mass clusters have already emptied their environment at $z>0.7$.

\section{Conclusions} \label{sec:conclusion}

We used our new AMASCFI cluster detection algorithm to detect 7100 cluster candidates with $S/N>3$ and $0.15<z\le1.1$ in the four Wide fields of the CFHTLS. Using lightcones extracted from the Millennium simulation, we derived the selection function of AMASCFI. At this $S/N$, we have a completeness of $\sim 50\%$ and a purity $> 80\%$ over the full redshift range and of $\sim 80\%$ and $\sim 90\%$ respectively for $z<0.7$. We computed a richness estimate for clusters with $S/N>4$ and $z<0.7$ and converted it to a mass estimate using a mass-richness scaling relation obtained from matching our cluster candidates with X-ray detected clusters. We obtained a catalogue of 1371 cluster candidates with mass $M_{200} > 10^{14} \ {\rm M_\odot}$, $S/N>4$ and $z<0.7$ in the four Wide fields of the CFHTLS, with a completeness of $\sim 70\%$ and purity $\sim 90\%$.

With our large and pure sample of cluster candidates, we are able to compute stacked GLFs for
ETGs, LTGs, and for the overall population in three redshift bins and
three cluster mass bins. Our main results are the follwing:

$\bullet$ the number of faint ETGs increases with decreasing redshift, the effect being stronger for higher mass clusters;

$\bullet$ the number of faint LTGs tends to be higher in higher mass clusters, whatever the redshift;

$\bullet$  the number density of faint LTGs increases only in low-mass clusters in the studied redshift range;

$\bullet$ the number density of bright LTGs decreases with decreasing redshift, whatever the mass.

These results show that the cluster red sequence is mainly formed at redshift $z>0.7$, and that faint ETGs continue to enrich the red sequence through quenching of brighter LTGs at $z\leq0.7$.  The efficiency of this quenching is higher in high-mass clusters, and the accretion rate of faint LTGs is higher in low-mass clusters as high-mass clusters have already accreted most of their environment.

The large number of clusters in our sample has allowed us to understand cluster galaxy evolution at intermediate redshifts ($0.15 \le z < 0.7$) and mass $M_{200} > 10^{14} \ M_{\rm{\sun}}$, and to break the degeneracy between these two observables. One interesting continuation would be to study the fraction of ETGs and LTGs in infalling filaments in order to help understand the accretion processes. Finally, the Euclid satellite will detect tens of thousands of clusters, allowing us to probe the higher redshift and lower mass clusters and groups.

\begin{acknowledgements}
We would like to thank the many people who invested in organizing and analyzing the {\it Euclid} Cluster Finder Challenge, in which AMASCFI was featured. This challenge certainly helped us in improving the performances of the code. We also thank the anonymous referee for useful remarks. F. Durret acknowledges long-term support from CNES. 
\end{acknowledgements}

%
%
\bibliographystyle{aa} 
\bibliography{GLFs_paper.bib} 

\appendix

\section{Red sequence versus \textit{LePhare} selection}

\begin{figure*}
         \centering 
          \includegraphics[width=1.0\textwidth]{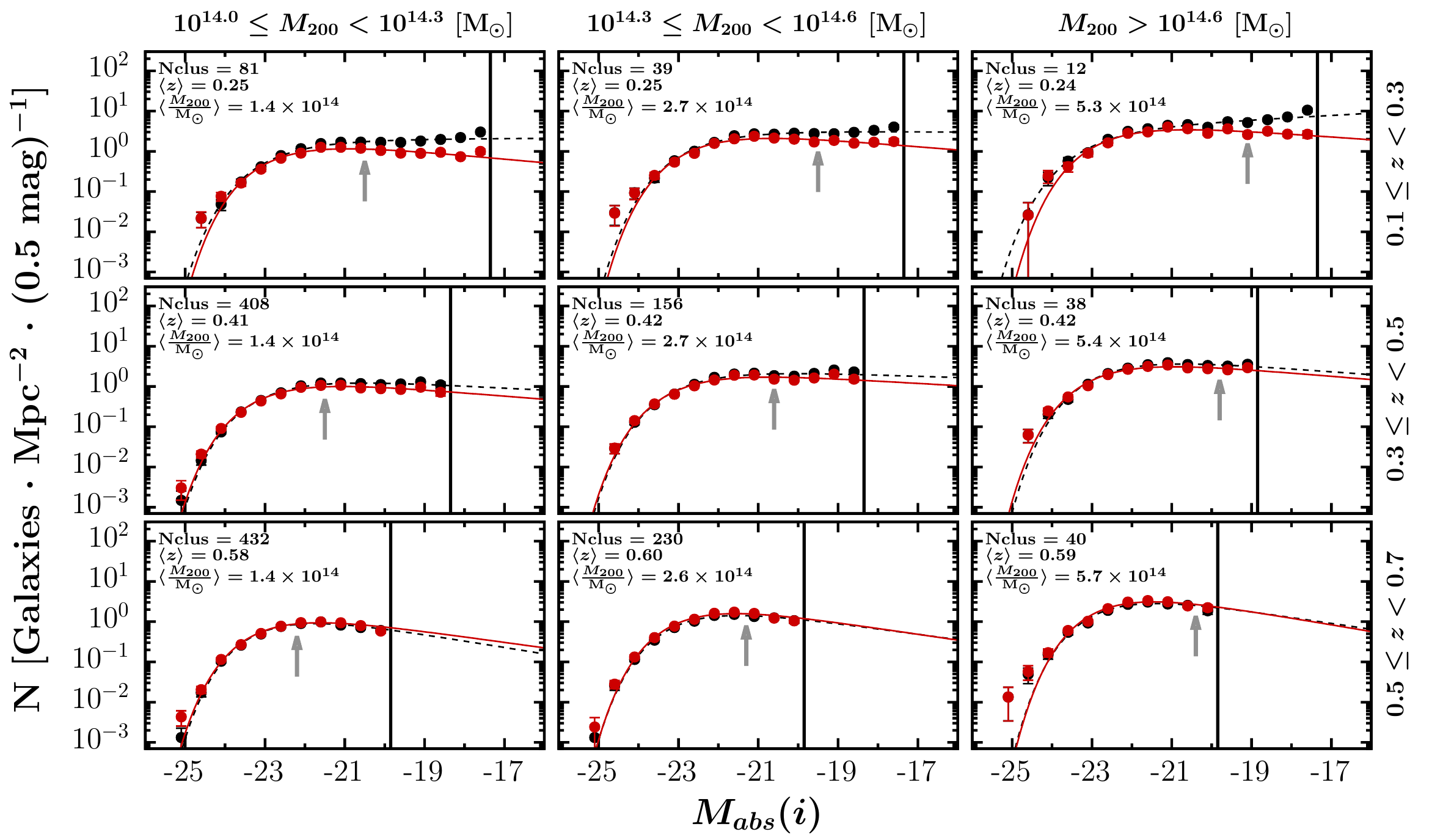}
     \caption{Redshift and mass co-evolution of the i-band stacked GLFs in the W1 field of the CFHTLS in bins of mass and redshift. In each panel are indicated the mean redshift and mass, and the number of cluster candidates in the bin. Black symbols are for the RS selection, while red symbols are for the \textit{LePhare} ETG selection. See Sect.~\ref{sect:ETG_LTG} for details.}
     \label{fig:stackGLF_zM-RS}
 \end{figure*}
   
\section{$\alpha$ versus $M^*$ confidence regions}\label{sect:alpha-vs_Mstar}

\begin{figure*}[!htb]
\centering
   \includegraphics[width=1.0\textwidth]{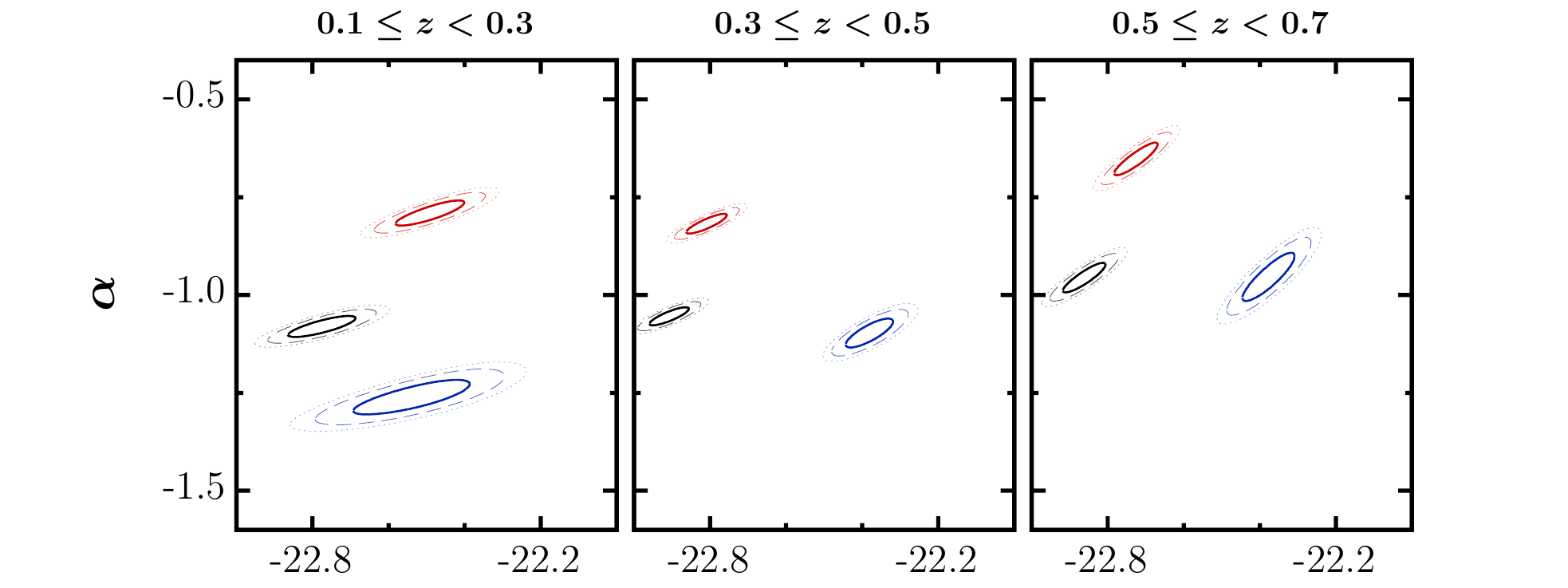}
     \caption{Contour plots of the 68\% (solid line), 95\% (dashed line), and 99\% (dotted line) confidence levels of the fit parameters for bins of redshift. Black is for all galaxies, red for ETGs, and blue for LTGs.}
     \label{fig:Mstar_vs_alpha-z}
   \end{figure*}
   
   \begin{figure*}[!htb]
\centering
   \includegraphics[width=1.0\textwidth]{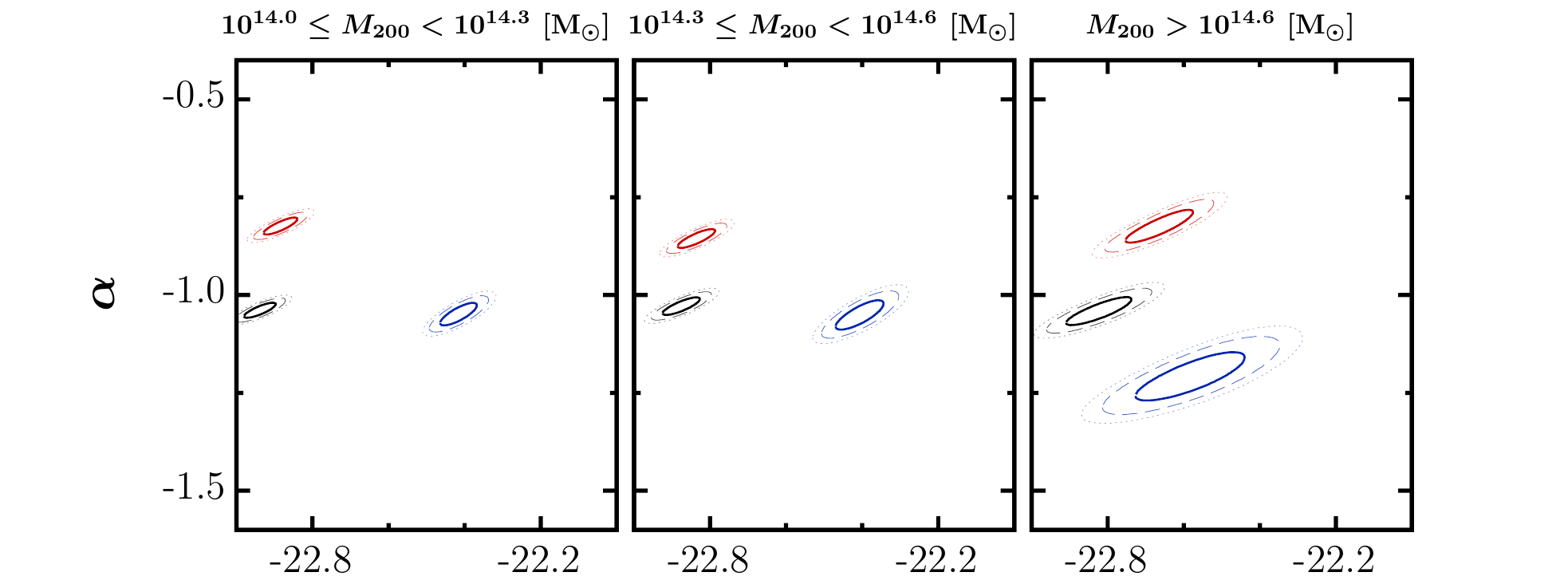}
     \caption{Contour plots of the 68\% (solid line), 95\% (dashed line), and 99\% (dotted line) confidence levels of the fit parameters for bins of mass. Black is for all galaxies, red for ETGs, and blue for LTGs.}
     \label{fig:Mstar_vs_alpha-M}
   \end{figure*}  

\begin{figure*}[!htb]
\centering
   \includegraphics[width=1.0\textwidth]{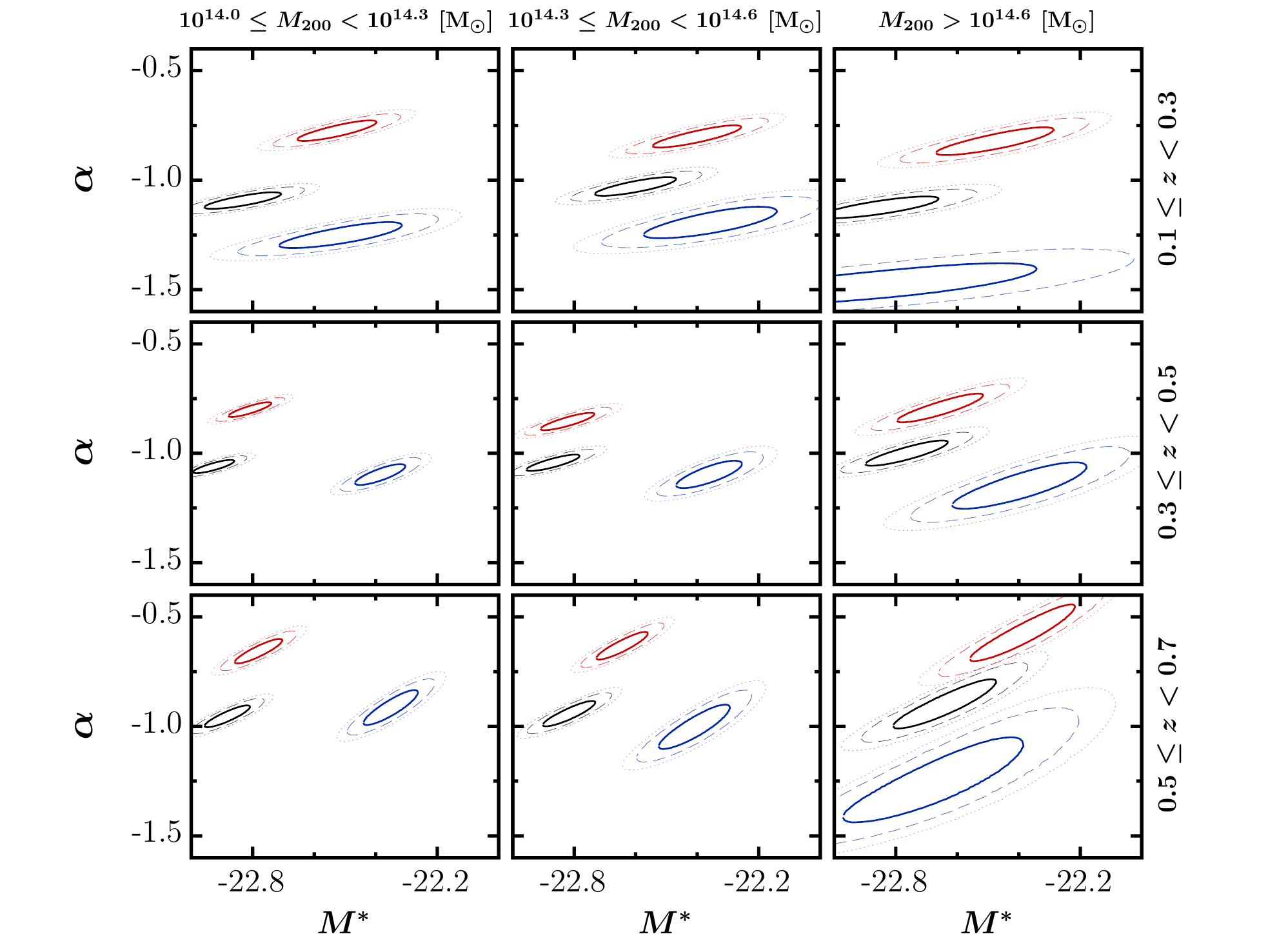}
     \caption{Contour plots of the 68\% (solid line), 95\% (dashed line), and 99\% (dotted line) confidence levels of the fit parameters for bins of redshift and mass. Black is for all galaxies, red for ETGs, and blue for LTGs.}
     \label{fig:Mstar_vs_alpha}
   \end{figure*}

\section{Supplementary tables}

\begin{landscape}
\begin{table}
\centering
\begin{tabular}{lc|ccccc}
\hline
\hline
type&$z$&$\alpha$ & $M^*$ & $\phi^*$ & $compl$ & $p$ \\
\hline
\hline
All &  0.2& -1.08$\pm$    0.02 &-22.78$\pm$ 0.06 &  2.9 $\pm$ 0.2 & -17.4 &   1.00 \\
&  0.4& -1.05$\pm$    0.02 &-22.91$\pm$ 0.03 &  2.8 $\pm$ 0.1 & -18.4 &   1.00 \\
&  0.6& -0.96$\pm$    0.02 &-22.86$\pm$ 0.04 & 3.7 $\pm$ 0.2 & -19.9 &   1.00 \\
\hline
ETGs &  0.2& -0.79$\pm$    0.02 &-22.49$\pm$ 0.06 &  3.0 $\pm$ 0.1 &  -17.4 &   1.00 \\
&  0.4& -0.82$\pm$    0.02 &-22.81$\pm$ 0.03 &  2.3 $\pm$ 0.1 & -18.4 &   1.00 \\
&  0.6& -0.65$\pm$    0.03 &-22.73$\pm$ 0.04 &  2.8 $\pm$ 0.1 & -19.9 &   1.00 \\
\hline
LTGs &  0.2& -1.26$\pm$    0.03 &-22.54$\pm$ 0.10 &  1.1 $\pm$ 0.1 &  -17.4 &   0.90 \\
&  0.4& -1.10$\pm$    0.02 &-22.38$\pm$ 0.04 &  3.0 $\pm$ 1.9 &  -18.4 &   1.00 \\
&  0.6& -0.95$\pm$    0.04 &-22.38$\pm$ 0.05 &  3.0 $\pm$ 2.7 &  -19.9 &   0.94 \\
\hline
\hline
\end{tabular}
\caption{Parameters of the Schechter fits to the stacked cluster GLFs in each redshift bin. As in Fig.~\ref{fig:stackGLF_z},  the central redshift of the slice is indicated. Galaxy clusters are selected in the range $z_{slice} \pm 0.1$. The `$compl$' and `$p$' columns correspond to the completeness limit to which the fit is done and the goodness of the fit normalized to one, respectively.}
\label{tab:schech_z}
\end{table}

\begin{table}
\centering
\begin{tabular}{lc|ccccc}
\hline
\hline
\hline
&$\mathrm{log}(\frac{M}{\rm M_{\odot}})$&$\alpha$ & $M^*$ & $\phi^*$ & $compl$ & $p$ \\
\hline
All & 14.15& -1.04$\pm$    0.01 &-22.94$\pm$ 0.03  &  2.5 $\pm$ 0.1 &  -17.4 &   1.00 \\
& 14.45& -1.03$\pm$    0.01 &-22.88$\pm$ 0.03 &  3.7 $\pm$ 0.1 &  -17.9 &   1.00 \\
& 14.75& -1.04$\pm$    0.02 &-22.82$\pm$ 0.06 &  6.0 $\pm$ 0.4 &  -18.4 &   1.00 \\
\hline
ETGs & 14.15& -0.82$\pm$    0.01 &-22.89$\pm$ 0.03 &  1.8 $\pm$ 0.1 &  -17.4 &   1.00 \\
& 14.45& -0.86$\pm$    0.02 &-22.84$\pm$ 0.03 &  2.9 $\pm$ 0.1 &  -17.9 &   1.00 \\
& 14.75& -0.83$\pm$    0.03 &-22.67$\pm$ 0.06 &  5.7 $\pm$ 0.3 &  -18.4 &   1.00 \\
\hline
LTGs & 14.15& -1.05$\pm$    0.02 &-22.42$\pm$ 0.03 &  1.9 $\pm$ 0.1 &  -17.4 &   1.00 \\
& 14.45& -1.05$\pm$    0.02 &-22.41$\pm$ 0.04 &  2.2 $\pm$ 0.1 &  -17.9 &   1.00 \\
& 14.75& -1.21$\pm$    0.04 &-22.58$\pm$ 0.09 &  2.0 $\pm$ 0.2 &  -18.4 &   0.72 \\
\hline
\hline
\end{tabular}
\caption{Parameters of the Schechter fits to the stacked cluster GLFs in each mass bin. The mass of the bin is indicated, `$compl$' and `$p$' correspond to the completeness limit to which the fit is done and the goodness of the fit normalized to one, respectively}
\label{tab:schech_M}
\end{table}

\begin{table}
\centering
\begin{tabular}{lc|ccc|ccc|ccc}
\hline
\hline
type& $z$ &  & M1 & & & M2 & & & M3\\
\hline
&&$\alpha$ & $M^*$ & $\phi^*$ &$\alpha$ & $M^*$ & $\phi^*$&$\alpha$& $M^*$ & $\phi^*$ \\
\hline
All  &   0.2 & -1.09$\pm$    0.02 &-22.83$\pm$  0.08 &  2.1$\pm$ 0.2 & -1.03$\pm$    0.03 &-22.60$\pm$  0.09 &  4.0$\pm$ 0.3 & -1.13$\pm$    0.03 &-22.86$\pm$  0.13 &  4.8$\pm$ 0.6 \\
&   0.4 & -1.06$\pm$ 0.02 &-22.93$\pm$  0.04 &  2.3$\pm$ 0.1 & -1.04$\pm$    0.02 &-22.87$\pm$  0.06 &  3.5$\pm$ 0.2 & -1.00$\pm$    0.04 &-22.76$\pm$  0.09 &  6.4$\pm$ 0.6 \\
&   0.6 & -0.95$\pm$ 0.03 &-22.88$\pm$  0.05 &  3.0$\pm$ 0.2 & -0.94$\pm$    0.04 &-22.82$\pm$  0.06 &  4.3$\pm$ 0.3 & -0.90$\pm$    0.07 &-22.64$\pm$  0.11 &  7.5$\pm$ 0.9 \\
\hline
ETGs  &   0.2 & -0.77$\pm$    0.03 &-22.53$\pm$  0.08 &  2.2$\pm$ 0.2 & -0.80$\pm$    0.03 &-22.40$\pm$  0.09 &  3.8$\pm$ 0.3 & -0.82$\pm$    0.04 &-22.48$\pm$  0.13 &  6.0$\pm$ 0.6 \\
&   0.4 & -0.80$\pm$ 0.02 &-22.81$\pm$  0.05 &  1.8$\pm$ 0.1 & -0.86$\pm$    0.03 &-22.82$\pm$  0.06 &  2.8$\pm$ 0.2 & -0.79$\pm$    0.04 &-22.66$\pm$  0.09 &  5.7$\pm$ 0.5 \\
&   0.6 & -0.66$\pm$ 0.04 &-22.78$\pm$  0.05 &  2.1$\pm$ 0.1 & -0.63$\pm$    0.04 &-22.65$\pm$  0.06 &  3.6$\pm$ 0.2 & -0.58$\pm$    0.09 &-22.39$\pm$  0.11 &  7.4$\pm$ 0.7 \\
\hline
LTGs  &   0.2 & -1.25$\pm$    0.04 &-22.51$\pm$  0.13 &  1.0$\pm$ 0.1 & -1.19$\pm$    0.05 &-22.36$\pm$  0.14 &  1.4$\pm$ 0.2 & -1.48$\pm$    0.07 &-22.91$\pm$  0.42 &  0.7$\pm$ 0.3 \\
&   0.4 & -1.10$\pm$ 0.03 &-22.39$\pm$  0.05 &  1.6$\pm$ 0.1 & -1.10$\pm$    0.04 &-22.36$\pm$  0.07 &  2.0$\pm$ 0.2 & -1.15$\pm$    0.07 &-22.40$\pm$  0.14 &  2.5$\pm$ 0.4 \\
&   0.6 & -0.92$\pm$ 0.05 &-22.35$\pm$  0.06 &  2.5$\pm$ 0.2 & -1.00$\pm$    0.07 &-22.41$\pm$  0.08 &  2.6$\pm$ 0.2 & -1.25$\pm$    0.13 &-22.67$\pm$  0.19 &  2.1$\pm$ 0.5 \\
\hline
\hline
\end{tabular}
\caption{Same as Table~\ref{tab:schech_z}, but for the three bins of mass.}
\label{tab:schech_zM}
\end{table}
\end{landscape}


\end{document}